\documentclass[prr,twocolumn,superscriptaddress]{revtex4}
\usepackage[utf8]{inputenc}
\usepackage[T1]{fontenc}
\usepackage{graphicx}
\usepackage{amsmath}
\usepackage{amssymb}
\usepackage{bm}
\usepackage{hyperref}
\usepackage{pgf}
\usepackage{comment}
\usepackage{chngcntr}
\usepackage[normalem]{ulem}

\newcommand*{\parallelogramm}
{ \rlap{\rotatebox{-30}{\rule[.05ex]{.4pt}{.77em}}}
  \kern.04em
  \rlap{\kern.36em\raisebox{0.649519052835em}{\rule{.6em}{.4pt}}}
  \rule{.6em}{.4pt}\kern-.04em
  \rotatebox{-30}{\rule[.05ex]{.4pt}{.77em}}}

\begin{document}

\title{Theory of charge-$6e$ condensed phase in Kagome lattice superconductors}
\author{Tong-Yu Lin}
\affiliation{State Key Laboratory of Low Dimensional Quantum Physics and Department of
Physics, Tsinghua University, Beijing 100084, China}
\author{Feng-Feng Song}
\affiliation{Institute for Solid State Physics, The University of Tokyo, Kashiwa, Chiba 277-8581, Japan}
\author{Guang-Ming Zhang}
\email{gmzhang@tsinghua.edu.cn}
\affiliation{State Key Laboratory of Low Dimensional Quantum Physics and Department of
Physics, Tsinghua University, Beijing 100084, China}
\affiliation{School of Physical Science and Technology, ShanghaiTech University, Shanghai,201210, China}
\affiliation{Frontier Science Center for Quantum Information, Beijing 100084, China}
\date{\today}

\begin{abstract}
We develop a Ginzburg-Landau theory for commensurate pair density wave (PDW) states in a hexagonal lattice system, relevant to the kagome superconductors $\rm{AV_3Sb_5}$. Compared to previous theoretical frameworks, the commensurate wave vectors permit additional symmetric terms in the free energy, altering the system's ground state and its degeneracy. In particular, we analyze topological defects in the energetically favorable $\psi_{\text{kagome}}$ ground state and find that kinks on domain walls can carry $1/3$ topological charges. We further establish a correspondence between the SC fluctuations in these states and an effective $J_1-J_2$ frustrated XY model on the emergent kagome lattice. By employing a state-of-the-art numerical tensor network method, we rigorously solve this effective model at finite temperatures and confirm the existence of a vestigial phase characterized by $1/3$ vortex-antivortex pairs in low temperatures with the absence of phase coherence of Cooper pairs, which is dual to the charge-$6e$ condensed phase. Our theory provides a potential explanation for the vestigial charge-$6e$ magnetoresistance oscillations observed in recent experiments [J. Ge, et. al., Phys. Rev. X 14, 021025 (2024)].
\end{abstract}

\maketitle

\section{Introduction.}
A pair-density wave (PDW) state is a superconducting (SC) state in which Cooper pairs carry a non-zero center-of-mass momentum \cite{Himeda_2002,Berg_2007,Berg_2009,Berg_2009_2,Agterberg_2008,Agterberg_2011,Agterberg_2020,Lee_2014,Wang_2015,Wang_2015_2,Jin_2022,Zhou_2022,Yao_2024,Pan_2022}. Phenomenologically, the PDW state breaks additional space group symmetries and is characterized by a SC order parameter with intricate spatial modulation. The interplay between the solid order and SC order in the PDW state can give rise to a rich spectrum of vestigial phases, including charge-density waves (CDWs), nematic order, and higher-charge SC.

The PDW state is believed to occur in a variety of materials, including cuprates \cite{Hamidian-2016,Ruan_2018,Du_2020,Edkins_2019,Li_2021,Gu_2023}, iron-based superconductors \cite{Liu_2023, Zhao_2023} and kagome superconducting materials. Among these, the kagome materials $\rm{AV_3Sb_5}$ have gained particular interest due to their exotic features in both normal states and SC states~\cite{Ortiz_2019,Ortiz_2020,Zhao_2021,Chen_2021,Tian_2023,Han_2024,Deng_2024}. $\rm{AV_3Sb_5}$ compounds crystallize in the hexagonal $P6/mmm$ space group with vanadium atoms forming a kagome network. The materials undergo CDW transitions at temperatures around $T_{\text{cdw}}\sim$ $80-110$K   \cite{Jiang_2021,Liang_2021,Li_2021,Ortiz_2021,Luo_2022,Hu_2022,Kang_2022,Nakayama_2022} and exhibit superconductivity at lower temperatures around $T_{\text{sc}} \sim 1$–$3$ K. In the SC state, a commensurate $3Q$ PDW order has been observed, where the SC order parameter oscillates in three distinct directions \cite{Chen_2021, Han_2024, Deng_2024}. This unusual PDW order is accompanied by evidence of time-reversal symmetry breaking (TRSB) in the SC phase \cite{Tian_2023}. Recently, magneto-transport experiments have revealed an exotic superconducting phase diagram in $\rm{CsV_3Sb_5}$~\cite{Ge_2023}. In the strongly fluctuating region above the conventional charge-$2e$ Cooper pair condensate, evidence of higher-charge superconductivity has emerged. Notably, magnetoresistance oscillations with a period of $\frac{h}{6e}$ were observed, indicating the existence of a charge-$6e$ condensate.

Despite the recent experimental progress, the physical origin of charge-$6e$ SC remains less understood. Several efforts have been made to explain this phenomenon. One prominent theoretical approach suggests an incommensurate hexagonal PDW state, where the $\psi_{\text{v-av}}$ states—featuring an emergent triangular lattice of vortices and anti-vortices—can lead to a vestigial charge-$6e$ superfluid phase through thermal melting \cite{Agterberg_2011}. More broadly, the interplay between composite order and phase coherence has become a valuable framework for understanding the complex phases of quantum materials. However, these existing theories cannot be directly applied to $\rm{AV_3Sb_5}$, as the latest experiments reveal a commensurate PDW state with a $2a_0 \times 2a_0$ period \cite{Chen_2021, Han_2024, Deng_2024}.

In this work, we propose a Ginzburg-Landau (GL) theory tailored for commensurate $2a_0 \times 2a_0$ PDW states in hexagonal lattice systems. Among the potential ground states, we focus on the energetically favorable $\psi_{\text{kagome}}$ state, where the order parameter breaks the time-reversal symmetry and forms an emergent vortex-antivortex kagome lattice. By analyzing phase coherence, we demonstrate that this state supports fractional vortex excitations, which manifest as kinks on the domain walls. Furthermore, we show that these fractional vortices play a crucial role in the system's thermal fluctuations, leading to the emergence of a charge-$6e$ coherent phase. The key aspect of phase coherence in the $\psi_{\text{kagome}}$ state can be effectively captured by a frustrated XY model on the kagome lattice, which shares the same ground-state structure and topological defects. This effective Hamiltonian offers a clear path for understanding the melting of orders from the perspective of the SC phase degrees of freedom, $\theta(r)$. By solving this model at finite temperatures using the tensor network method, we confirm the existence of a vestigial charge-$6e$ coherent phase above the charge-$2e$ condensate, characterized by power-law correlations in $e^{i3\theta}$ and exponential decay in $e^{i\theta}$. Finally, we discuss the experimental implications of our findings and how they may guide future investigations.

\section{Ginzburg-Landau theory}

\begin{figure}[t]
\centering
\includegraphics[width=\linewidth]{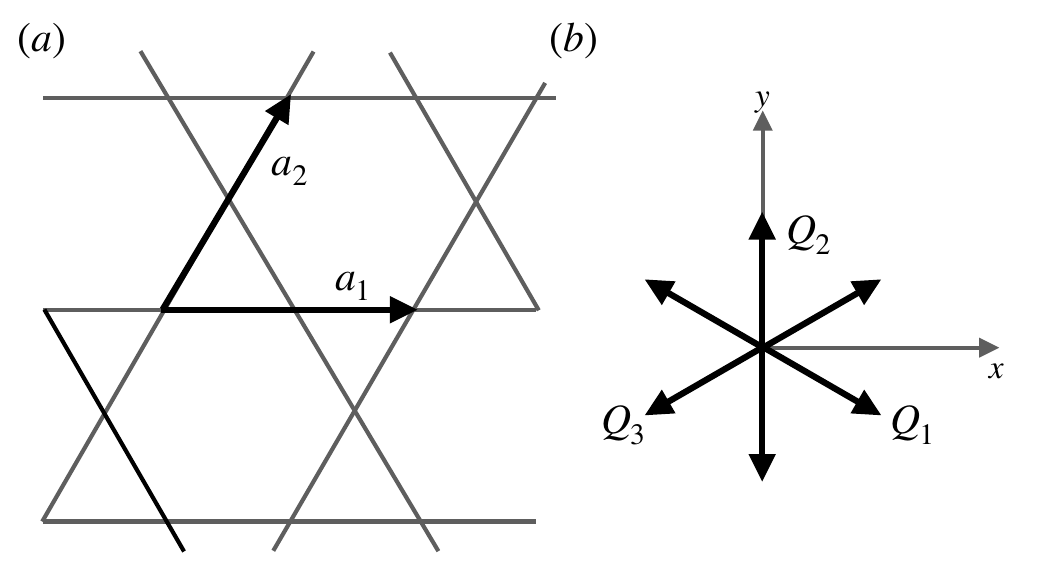}
\caption{(a) Kagome lattice. (b) Directions of wave vectors $\bm{Q}_i$.}
\label{fig:kagome_lattice}
\end{figure}

We start with the six-component order parameter for the 3Q PDW state as follows:
\begin{equation}
\Delta_{\text{pdw}}(r)=\sum_{j=1,2,3}\Delta_{\bm{Q}_j}e^{i\bm{Q}_j\cdot \bm{r}}+\sum_{j=1,2,3}\Delta_{-\bm{Q}_j}e^{-i\bm{Q}_j\cdot \bm{r}}.
\end{equation}
As shown in Fig.~\ref{fig:kagome_lattice}(a)-(b), $\bm{Q}_1$ and $\bm{Q}_2$ are chosen to be half of the Bragg vectors of the underlying kagome lattice, with $\bm{Q}_3 = -\bm{Q}_1 - \bm{Q}_2$. Specifically, these wave vectors are given by $\bm{Q}_1 = \frac{\pi}{|a_1|}(1, -\frac{1}{\sqrt{3}})$ and $\bm{Q}_2 = \frac{\pi}{|a_2|}(0, \frac{2}{\sqrt{3}})$, where $a_1$ and $a_2$ are the lattice constants. This choice of wavevectors aligns with evidence for the $2a_0 \times 2a_0$ 3Q pair-density wave (PDW) state observed in recent experiments~\cite{Han_2024, Deng_2024}.

The GL free energy density is constructed by imposing the symmetries of the space group $P6/mmm$, $U(1)$ gauge symmetry, and time-reversal symmetry. For commensurate wave vectors with a periodicity of $2a_0$, the GL free energy density up to fourth order is given by:
\begin{equation}
F(\{\Delta_{\bm{Q}}\})=F_I(\{\Delta_{\bm{Q}}\})+F_c^{(2)}(\{\Delta_{\bm{Q}}\})+F_c^{(4)}(\{\Delta_{\bm{Q}}\}),
\end{equation}
where $F_I$ is the GL free energy density without considering the underlying lattice~\cite{Agterberg_2011}
\begin{widetext}
\begin{align}
\begin{split}
F_I=&-\alpha\sum_{i}|\Delta_{\bm{Q}_i}|^2+\beta_1(\sum_{i}|\Delta_{\bm{Q}_i}|^2)^2+
\beta_2\sum_{i}|\Delta_{\bm{Q}_i}|^2|\Delta_{-\bm{Q}_i}|^2\\
&+\beta_3(|\Delta_{\bm{Q}_1}|^2|\Delta_{\bm{Q}_2}|^2+|\Delta_{\bm{Q}_1}|^2|\Delta_{\bm{Q}_3}|^2+|\Delta_{\bm{Q}_2}|^2|\Delta_{\bm{Q}_3}|^2+|\Delta_{-\bm{Q}_1}|^2|\Delta_{-\bm{Q}_2}|^2+|\Delta_{-\bm{Q}_1}|^2|\Delta_{-\bm{Q}_3}|^2+|\Delta_{-\bm{Q}_2}|^2|\Delta_{-\bm{Q}_3}|^2)\\
&+\beta_4(|\Delta_{\bm{Q}_1}|^2|\Delta_{-\bm{Q}_2}|^2
+|\Delta_{\bm{Q}_1}|^2|\Delta_{-\bm{Q}_3}|^2+|\Delta_{\bm{Q}_2}|^2|\Delta_{-\bm{Q}_3}|^2+|\Delta_{-\bm{Q}_1}|^2|\Delta_{\bm{Q}_2}|^2+|\Delta_{-\bm{Q}_1}|^2|\Delta_{\bm{Q}_3}|^2+|\Delta_{-\bm{Q}_2}|^2|\Delta_{\bm{Q}_3}|^2)\\
&+\beta_5[\Delta_{\bm{Q}_1}\Delta_{-\bm{Q}_1}(\Delta_{\bm{Q}_2}\Delta_{-\bm{Q}_2})^*+\Delta_{\bm{Q}_1}\Delta_{-\bm{Q}_1}(\Delta_{\bm{Q}_3}\Delta_{-\bm{Q}_3})^*+\Delta_{\bm{Q}_2}\Delta_{-\bm{Q}_2}(\Delta_{\bm{Q}_3}\Delta_{-\bm{Q}_3})^*
+c.c]+\kappa_1\sum_i|\nabla \Delta_{\bm{Q}_i}|^2\\
&+ \kappa_2[(|\nabla_+\Delta_{\bm{Q}_1}|^2+|\nabla_+\Delta_{-\bm{Q}_1}|^2)+ e^{i2\pi/3}(|\nabla_+\Delta_{\bm{Q}_2}|^2+|\nabla_+\Delta_{-\bm{Q}_2}|^2)+ e^{i4\pi/3}(|\nabla_+\Delta_{\bm{Q}_3}|^2+|\nabla_+\Delta_{-\bm{Q}_3}|^2)+c.c.]
\end{split}
\label{eqn: incommensurate free energy}
\end{align}
with $c.c$ denoting the complex conjugate, $\nabla_{\pm}=\nabla_x\pm i\nabla_y$. The second- and fourth-order commensurate terms, $F_c^{(2)}$ and $F_c^{(4)}$, specifically introduced to account for the $2a_0 \times 2a_0$ PDW, are given by
\begin{align}
    F_c^{(2)}=&\gamma\sum_{j=1,2,3}\Delta_{\bm{Q}_j}\Delta_{-\bm{Q}_j}^{*}+c.c;  \label{eqn: second order}\\
     F_c^{(4)}=&\gamma_1\sum_{j=1,2,3}|\Delta_{\bm{Q}_j}|^2(\sum_{j=1,2,3}\Delta_{\bm{Q}_j}\Delta_{-\bm{Q}_j}^{*}+c.c)
    +\gamma_2(\Delta_{\bm{Q}_1}\Delta_{-\bm{Q}_1}^{*}\Delta_{\bm{Q}_2}\Delta_{-\bm{Q}_2}^{*}+\\ \notag
    &\qquad\Delta_{\bm{Q}_1}\Delta_{-\bm{Q}_1}^{*}\Delta_{\bm{Q}_3}\Delta_{-\bm{Q}_3}^{*}
    +\Delta_{\bm{Q}_2}\Delta_{-\bm{Q}_2}^{*}\Delta_{\bm{Q}_3}\Delta_{-\bm{Q}_3}^{*}+c.c). \label{eqn: Fourth order}
\end{align}
\end{widetext}

To ensure the stability of the free energy, we require that $\alpha>0$ and $\beta_1>0$. Furthermore, since $F_c^{(2)}$ is a second-order term, we assume it dominates the GL free energy. The amplitude of $\Delta_{\bm{Q}_i}$ is determined by the terms with coefficients $\{\alpha,\beta_1,\beta_2,\beta_3,\beta_4\}$, while the relative phase of each component $\{\Delta_{\bm{Q}_i}\}$ is governed by the terms with coefficients $\{\beta_5,\gamma,\gamma_1,\gamma_2\}$.

\begin{table*}[t]
\caption{Possible commensurate $3Q$ PDW ground states, where the relative phases of each PDW component are locked into $Z_2$ values, given by $\phi_i = 0$ or $\pi$. G.S.D. stands for ground state degeneracy.}
\begin{tabular}{|c|c|c|c|c|c|}
  \hline
 &State & $\gamma<0$ & State & $\gamma>0$& G.S.D \\ \hline
$\beta_5<0$&$\Psi_{\triangle}$&$\Delta e^{i\theta}\sum_{j=1,2,3}\cos [\bm{Q}_j\cdot \bm{r}+\phi_j]$&$\Psi_{hc}$&$\Delta e^{i\theta} \sum_{j=1,2,3}
\sin [\bm{Q}_j\cdot \bm{r}+\phi_j]$& $U(1)\times Z_2\times Z_2$ \\ \hline
 $\beta_5>0$ &$\Psi_{kag}$&$\Delta e^{i\theta}\sum_{j=1,2,3} e^{\pm i(j-2)\frac{\pi}{3}} \cos[\bm{Q}_j\cdot \bm{r}+\phi_j]$&$\Psi_{Ruby}$&$\Delta e^{i\theta}\sum_{j=1,2,3} e^{\pm i(j-2)\frac{\pi}{3}} \sin[\bm{Q}_j\cdot \bm{r}+\phi_j]$ & $U(1)\times Z_2\times Z_2\times Z_2$ \\  \hline
\end{tabular}
\label{tab1}
\end{table*}

Based on STM results from \cite{Deng_2024}, we consider the possible ground states of the system where all six paring components are non-zero. When $\beta_2<0,\beta_3<0,\beta_4<0$ in Eq.\eqref{eqn: incommensurate free energy}, the states with $|\Delta_{\bm{Q}_1}|=|\Delta_{\bm{Q}_2}|=|\Delta_{\bm{Q}_3}|=|\Delta_{-\bm{Q}_1}|=|\Delta_{-\bm{Q}_2}|=|\Delta_{-\bm{Q}_3}|\neq 0$ are energetically favored. Depending on the signs of $\{\beta_5,\gamma\}$, there are four possible ground states, as listed in Table~\ref{tab1}. Here, $\Delta e^{i\theta}$ denotes the overall amplitude and phase of the PDW state, while $\{\phi_i\}$ determines the center of the PDW. The phase of each PDW component is locked to a $Z_2$ value, as dictated by Eq.~\eqref{eqn: second order}. Moreover, the state with $(\phi_1,\phi_2,\phi_3)=(0,0,0)$ can be transformed into the state with $(\phi_1,\phi_2,\phi_3)=(\pi,\pi,\pi)$ by the transformation $\Delta\rightarrow -\Delta$. Therefore, the variables $\{\phi_i\}$ lead to a $Z_2\times Z_2$ ground state degeneracy, corresponding to the breaking of translational symmetry.

For $\beta_5<0$, in the $\psi_{\triangle}$ state, the positions of the maximum amplitude form a triangular lattice, while in the $\psi_{hc}$ state, the positions of the maximum amplitude form a honeycomb lattice. The phase distribution in both states is uniform and there are no SC vortices in the ground state. Since neither state breaks time-reversal symmetry (TRS), they do not require further consideration.

For $\beta_5>0$, the ground state with time-reversal symmetry breaking is energetically preferred. There are two possible ground states, the $\psi_{\text{kagome}}$ state and the $\psi_{\text{Ruby}}$ state. The amplitude distribution of the SC order parameters $\psi_{\text{kagome}}(r)$ is illustrated in Fig.\ref{fig:kagome_ground_state}(a), whose maximum forms an emergent kagome lattice. Zeros in the amplitude, which correspond to the positions of vortices and antivortices, are found at the centers of triangular and hexagonal plaquettes. Specifically, $\pm 1$ vortices are located at the centers of the triangles, while $\mp 2$ vortices reside at the centers of the hexagons, as shown in Fig.\ref{fig:kagome_ground_state}(b). The $\psi_{\text{Ruby}}(r)$ state features a similar vortex structure, with the maxima forming a Ruby lattice, as detailed in Appendix~\ref{sec:phi_ruby}.

\section{Fractional vortex excitations}
At low temperatures, the amplitude fluctuations of the SC order parameter can be ignored, leaving the thermal fluctuations of the system governed primarily by phase variations. These phase fluctuations include the global SC phase $\theta$ and the relative phases $\{\phi_i\}$ of each component. The phase $\theta$ describes the overall SC degree of freedom, with $\{\phi_i\}$ corresponding to the two phonon modes of the emergent PDW lattice, where $\bm{u} = (u_x, u_y)$ and $\phi_i = \bm{Q}_i \cdot \bm{u}$. Since the $F_c^{(2)}$ term dominates, the commensurate locking effect is strong, effectively freezing the phonon modes at low temperatures. Consequently, the thermal fluctuations are primarily driven by variations in the overall SC phase $\theta$.

The phase fluctuations can be analyzed by classifying the topological defects in the order parameter. With the $Z_2 \times Z_2$ degeneracy associated with $\{\phi_i\}$ frozen, the low-energy topological defects are governed by the $U(1) \times Z_2$ degeneracy in both the $\psi_{\text{kagome}}$ and $\psi_{\text{Ruby}}$ states. Due to the presence of $U(1)$ symmetry, only quasi-long-range order (QLRO) in the $\theta$ field can be maintained at low temperatures, even in the presence of long-range PDW patterns, in accordance with the Mermin-Wagner theorem. The additional $Z_2$ degeneracy, arising from time-reversal symmetry breaking (TRSB), allows the flipping of all vortex signs. Since both $\psi_{\text{kagome}}$ and $\psi_{\text{Ruby}}$ states share the same degeneracy, their topological defects can be constructed in a similar manner. In the following, we focus on the $\psi_{\text{kagome}}$ state, while a similar analysis for the $\psi_{\text{Ruby}}$ state is provided in Appendix~\ref{sec:phi_ruby}.

\begin{figure}[t]
\centering
\includegraphics[width=\linewidth]{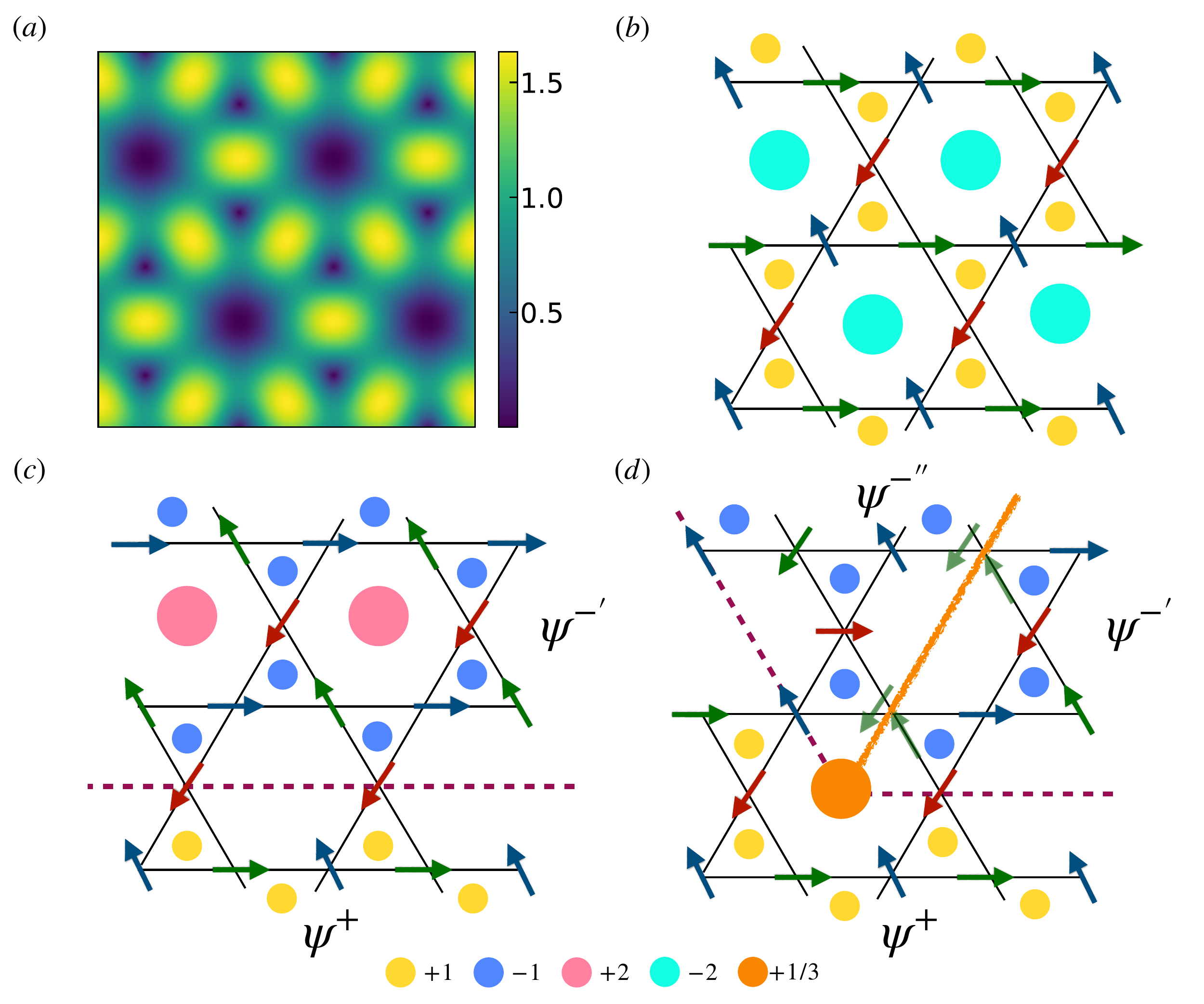}
\caption{(a) The amplitude distribution of $\protect\psi_{\text{kagome}}$. (b) The phase distribution of the $\protect\psi_{\text{kagome}}$ state on the emergent kagome lattice, which also describes the ground state of the AFXY model \eqref{eqn:hamiltonian}. The circles in the plaquette denote vortex distribution in the $\protect\psi_{\text{kagome}}$ state. (c) Order parameter with infinite straight domain wall, depicted by the purple dashed line. (d) The $+\frac{1}{3}$ vortex excitation at the kink of the domain wall.}
\label{fig:kagome_ground_state}
\end{figure}

In line with the ground state degeneracy, two types of topological excitations arise: vortices and domain walls. We first consider an infinite straight domain wall in the $\psi_{\text{kagome}}$ state located at the following positions:
\begin{equation}
    \bm{Q}_j\cdot \bm{r}=n\pi,\quad n\in Z.
\end{equation}
As shown in Fig.\ref{fig:kagome_ground_state} (c), the domain wall goes through the centers of hexagons of the emergent kagome lattice. When the order parameter on one side of the wall is fixed as
\begin{equation}
    \psi^{+}(r)=\Delta\sum_{j=1,2,3} e^{+i(j-2)\frac{\pi}{3}}\cos(\bm{Q}_j\cdot \bm{r}+\phi_j). \label{eqn: domain wall excitation 1}
\end{equation}
The energy of the system is minimized when the order parameters on the other side of the wall are given by:
\begin{equation}
\psi^{-'}(r)=\Delta e^{i\frac{2\pi}{3}}\sum_{j=1,2,3} e^{-i(j-2)\frac{\pi}{3}}\cos(\bm{Q}_j\cdot \bm{r}+\phi_j), \label{eqn: domain wall excitation 2}
\end{equation}
ensuring that the order parameters in Eqs.~\eqref{eqn: domain wall excitation 1} and \eqref{eqn: domain wall excitation 2} remain continuous across the domain wall.

Another important type of topological defect is the $1/3$ fractional vortex, which appears as a kink on a domain wall. Consider two half-infinite domain walls defined by:
\begin{equation}
    \bm{Q}_2 \cdot \bm{r} = n\pi, \quad \bm{Q}_3 \cdot \bm{r} = m\pi.
\end{equation}
As shown in Fig.~\ref{fig:kagome_ground_state}(d), two domain walls intersect at the center of a hexagonal plaquette, forming a kink. To locally minimize the energy, three distinct order parameters arise which are separated by two domain walls: $\psi_{\text{kagome}}^{+}(r)$, $\psi^{-'}(r)$ and
\begin{equation}
\psi^{-''}(r)=\Delta \sum_{j=1,2,3} e^{-i(j-2)\frac{\pi}{3}}\cos(\bm{Q}_j\cdot \bm{r}+\phi_j).
\label{eqn: domain wall excitation 3}
\end{equation}
When going across two domain walls, a global phase discrepancy of $\frac{2\pi}{3}$ accumulates, characterized by a semi-infinite branch cut starting from the center of the hexagon, as depicted by the orange line in Fig.\ref{fig:kagome_ground_state}(d). To compensate for this phase mismatch, the phase of the order parameters around the kink has to rotate by an angle of $\frac{2\pi}{3}$, effectively forming a $\frac{1}{3}$ fractional vortex. These fractional topological excitations are crucial in the understanding of the melting of the PDW phases.

\section{Effective model}

To capture the most essential physics of the topological excitations in the GL theory, we consider an effective XY spin model with spin orientations corresponding to the phase angles of the SC order parameter. Specifically, we consider the $J_1$-$J_2$ antiferromagnetic XY (AFXY) model on the emergent kagome lattice \cite{Huse_1992, Korshunov_2002}, defined by the Hamiltonian:
\begin{equation}
    H_{\text{kagome}} = J_1 \sum_{\langle i,j \rangle} \cos(\theta_i - \theta_j) + J_2 \sum_{\langle\langle i,j \rangle\rangle} \cos(\theta_i - \theta_j),
    \label{eqn:hamiltonian}
\end{equation}
where $J_1$ and $J_2$ represent antiferromagnetic interactions between nearest neighbors (NN) and next-to-nearest neighbors (NNN), respectively. The reason is that the ground state of the AFXY kagome model perfectly matches that of the vortex-antivortex lattice in the commensurate PDW state, along with its topological excitations. Moreover, the ``kink in the domain wall'' mechanism in such frustrated XY systems~\cite{Korshunov_2006} provides valuable insight into the melting of the PDW order.

As shown in Fig.~\ref{fig:kagome_ground_state}(b), the ground-state spin configurations of the AFXY kagome model \eqref{eqn:hamiltonian} align with the phase distribution of the $\psi_{\text{kagome}}$ states on the emergent kagome lattice. These configurations are characterized by the spin orientations on three sublattices, labeled $(\theta_A, \theta_B, \theta_C)$, with an angular difference of $2\pi/3$ between each other. The vortex distribution in the PDW states can be described by the vorticity \cite{Vallat_1994}, defined as
\begin{equation}
v_p = \frac{1}{2\pi} \sum_{i,j \in p} \phi_{ij}, \quad \phi_{ij} = (\theta_i - \theta_j) \in[-\pi,\pi),
\end{equation}
where the summation is taken clockwise over the bonds around the plaquette $p$. It can be easily verified that the vorticity in the lattice model is consistent with that of the PDW states. Furthermore, the ground state of the lattice model exhibits a $U(1) \times Z_2$ degeneracy and shares the same class of topological defects as the PDW states \cite{Korshunov_2002}, as illustrated in Fig.~\ref{fig:kagome_ground_state}(c)-(d).

After establishing the connection between thermal fluctuations in the $\psi_{\text{kagome}}$ state and $J_1-J_2$ AFXY model, we now turn to a discussion of the potential phase transitions in the effective model, with a particular focus on the physically relevant regime where $J_2\ll J_1$. Notably, since fractional vortices manifest as kinks along domain walls, they cannot dissociate below the temperature at which domain walls proliferate. These fractional vortices are bound not only by their logarithmic interaction but also by the energy cost of the domain walls connecting them. As a result, the temperature for fractional vortex unbinding, $T_{\text{fv}}$, must satisfy $T_{\text{dw}}\le T_{\text{fv}}$, where $T_{\text{dw}}$ is the temperature for domain wall proliferation.

Whether the two transitions occur at the same temperature depends on the ratio $J_2/J_1$. The energy cost of a domain wall per unit length can be estimated as $E_{\text{dw}} \propto J_2$, which is primarily determined by NNN interactions, while the energy cost of fractional vortices is mainly governed by NN interactions, $E_{\text{fv}} \propto J_1$. The entropy per unit length of the domain wall, $S_{\text{dw}}$, is influenced by the possibility of kinks forming along it and can be approximated as $S_{\text{dw}} \sim \exp(-E_{\text{fv}}/T)$. A direct comparison between the energy and entropy contributions, through the free energy expression $F_{\text{dw}} = E_{\text{dw}} - T S_{\text{dw}}=0$, reveals an upper bound on the ratio $J_2/J_1 = J^*$. Below this threshold, the two types of topological defects can proliferate independently, with $T_{\text{dw}} < T_{\text{fv}}$ \cite{Korshunov_2002}.

In the temperature range between $T_{\text{dw}}$ and $T_{\text{fv}}$, phase coherence between integer vortices is destroyed due to the proliferation of domain walls. However, the logarithmic interaction between fractional vortices remains strong enough to bind them into pairs, resulting in a fractional vortex-paired phase characterized by QLRO in the field $e^{i3\theta(r)}$~\cite{Huse_1992,Korshunov_2002}. This phase eventually melts via a Berezinskii-Kosterlitz-Thouless (BKT) transition, driven by the unbinding of $\frac{1}{3}$ vortices. For the $\psi_{\text{kagome}}$ states, the field $e^{i\theta(r)}$ corresponds to the order parameter of charge-$2e $ PDW order parameter, given by: $\Delta_{\bm{Q}_i}\rightarrow e^{i\theta}\Delta_{\bm{Q}_i}$. Correspondingly, the field $e^{i3\theta(r)}$ represents the phase of the charge-$6e$ order parameter in the pair-density wave (PDW) state, expressed as $\langle \Delta_{\bm{Q}_1} \Delta_{\bm{Q}_2} \Delta_{\bm{Q}_3} \rangle$. Therefore, the fractional vortex-paired phase can be interpreted as a charge-$6e$ ordered coherent phase.

A direct consequence of QLRO in the field $e^{i3\theta(r)}$ is the flux quantization of $\frac{h}{6e}$. If an external magnetic field is applied, the superconducting state couples to the gauge field through minimal coupling: $H \sim |(-i\hbar \nabla - 2eA)\psi|^2$. When the magnetic flux equals $\frac{1}{3}$ flux quanta, $\frac{1}{3}$ vortices are excited, inducing a phase gradient $\int \nabla \theta \cdot dl = \frac{2e}{\hbar}\int A \cdot dl = \frac{2\pi}{3}$, which minimizes the system's free energy. This results in the oscillatory behavior with a period of $\frac{h}{6e}$. Consequently, the phase transition between the $\psi_{\text{kagome}}$ states and the charge-$6e$ coherent phase offers a plausible explanation for the flux quantization observed in experiments~\cite{Ge_2023}.

\section{Numerical results}

From the previous analysis, the phase transition temperature $T_{\text{dw}}$ is estimated to be lower than the BKT transition temperature $T_{\text{fv}} \sim 10^{-2}J_1$, driven by the unbinding of $\frac{1}{3}$ vortices. However, due to the frustration effects~\cite{Rakala_2017} and critical slowing down at low temperatures~\cite{Swendsen_1987,Wolff_1989}, this temperature regime is challenging to access using standard Monte Carlo methods~\cite{Kakizawa_2023}. To address this difficulty, we utilize a recently developed tensor network method tailored for frustrated spin models~\cite{Song_2023_2}. For completeness, we provide a brief outline of the method here, with details available in Appendix~\ref{sec: tensor}.

The partition function of AFXY kagome model is expressed as:
\begin{equation}
    Z=\prod_{i}\int\frac{d\theta_i}{2\pi}\exp[-\beta H_{\text{kagome}}(\{\theta_i\})].
    \label{eqn:partition function}
\end{equation}
In the tensor network (TN) formalism, the 2D partition function is represented as an infinite 2D tensor network. This can be further transformed into a 1D quantum transfer matrix in the form of a matrix product operator. In the thermodynamic limit, solving the partition function reduces to finding the leading eigenvalue and corresponding eigenvectors of the transfer matrix, which can be accurately obtained using the multi-site variational uniform matrix product state (VUMPS) algorithms \cite{Nietner_2020}. With the leading eigenvectors, various physical quantities, including local observables and correlation functions, can be efficiently computed. Additionally, since the classical partition function is mapping into a quantum model, the entanglement measure of quantum systems can be used as a precise criterion for identifying phase transitions.

The quantum transfer matrix of Eq.\eqref{eqn:partition function} is non-hermitian and
we employ the generalized entanglement entropy \cite%
{Yi-Ting_2022,Tang_2023} to determine the phase boundary. In Fig.\ref%
{fig:EE_result}(a), the numerical results of the generalized entanglement
entropy at $J_{2}=0.0025J_{1}$, with bond dimensions ranging from $D=40$ to $%
D=120$ are displayed. The entanglement entropy exhibits a sharp peak and a
discontinuous jump, corresponding to continuous and discontinuity phase
transitions, respectively. The precise transition temperatures can be
determined by extrapolating the bond dimension. As shown in Fig.~\ref%
{fig:EE_result}(b), the lower transition temperature $T_{c1}\simeq 0.035J_{1}
$ remains nearly unchanged as the bond dimension increases, while the higher
transition temperature is $T_{c2}\simeq 0.072J_{1}$, which is close to the
estimated BKT transition temperature driven by $\frac{1}{3}$ vortex: $%
T_{fv}\approx \frac{1}{9}T_{v}\approx 0.075J_{1}$.

\begin{figure}[t]
\centering
\includegraphics[width=\linewidth]{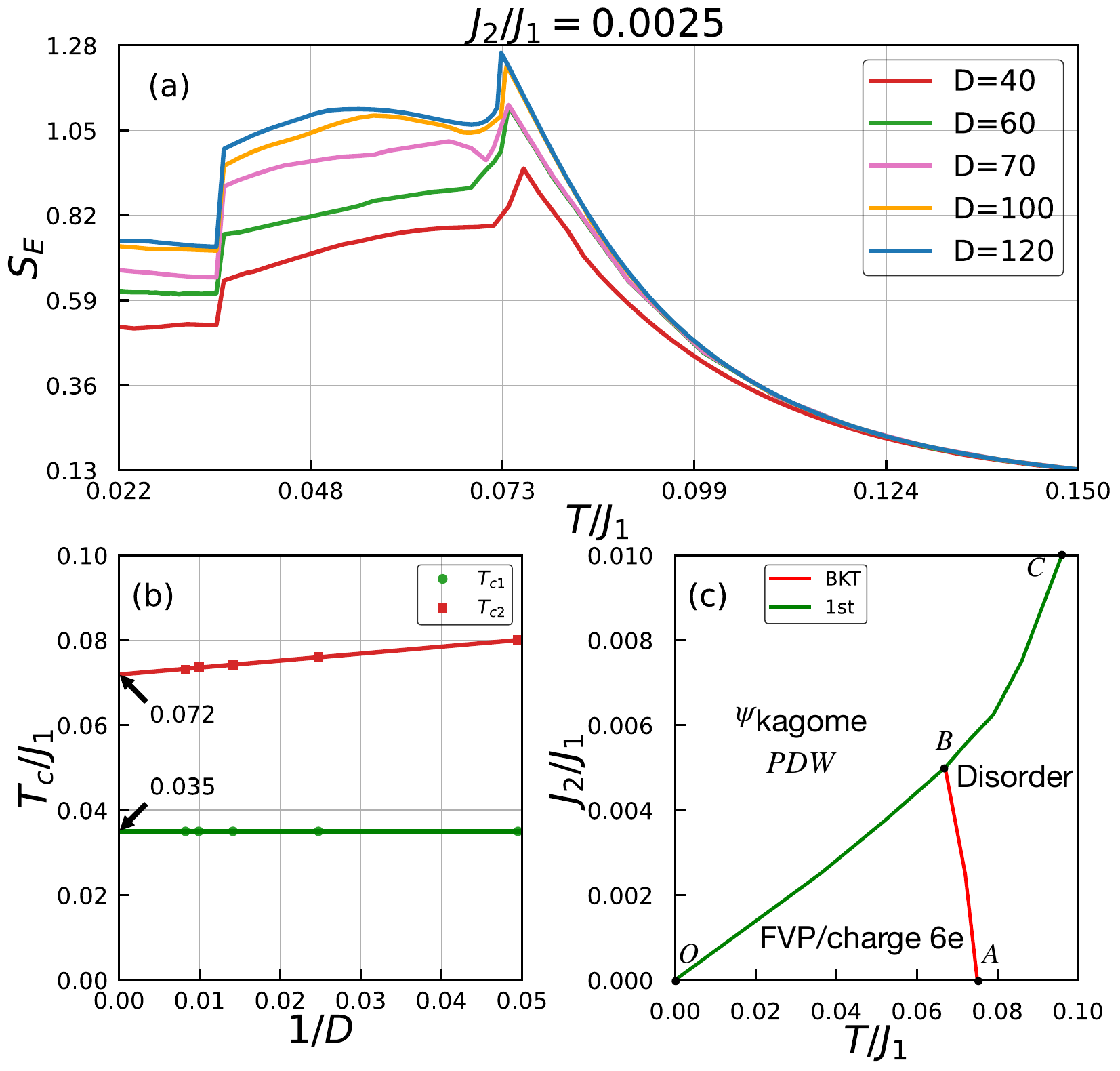}
\caption{(a) The generalized entanglement entropy as a function of
temperature for $J_2=0.0025J_1$. (b) The transition temperatures $T_{c1}$
and $T_{c2}$ of the entanglement entropy extrapolated from MPS bond dimensions from
$40$ to $120$. (c) The phase diagram of the $J_1-J_2$ AF XY spin model on a
kagome lattice. The BKT transition line starting from point A (0.075,0) and
terminating at point B. The coordinates of point B are estimated as $%
J_2\approx 0.005\sim 0.0055J_1$, $T\approx 0.067J_1\sim 0.070J_1$.}
\label{fig:EE_result}
\end{figure}

For $J_2>0.0055J_1$, the generalized entanglement entropy exhibits one
discontinuous jump only. Then a phase diagram can be calculated and shown in
Fig.\ref{fig:EE_result}(c), where the low-temperature PDW phase
is characterized by QLRO in $\langle e^{i\theta}\rangle$ and long-range
order (LRO) in vorticity variables. When crossing the first-order
transition, the chiral LRO and phase QLRO break simultaneously, indicating the melting of the vortex lattice. The
extended numerical data are shown in Appendix \ref{sec: data}.

\begin{figure}[t]
\centering
\includegraphics[width=\linewidth]{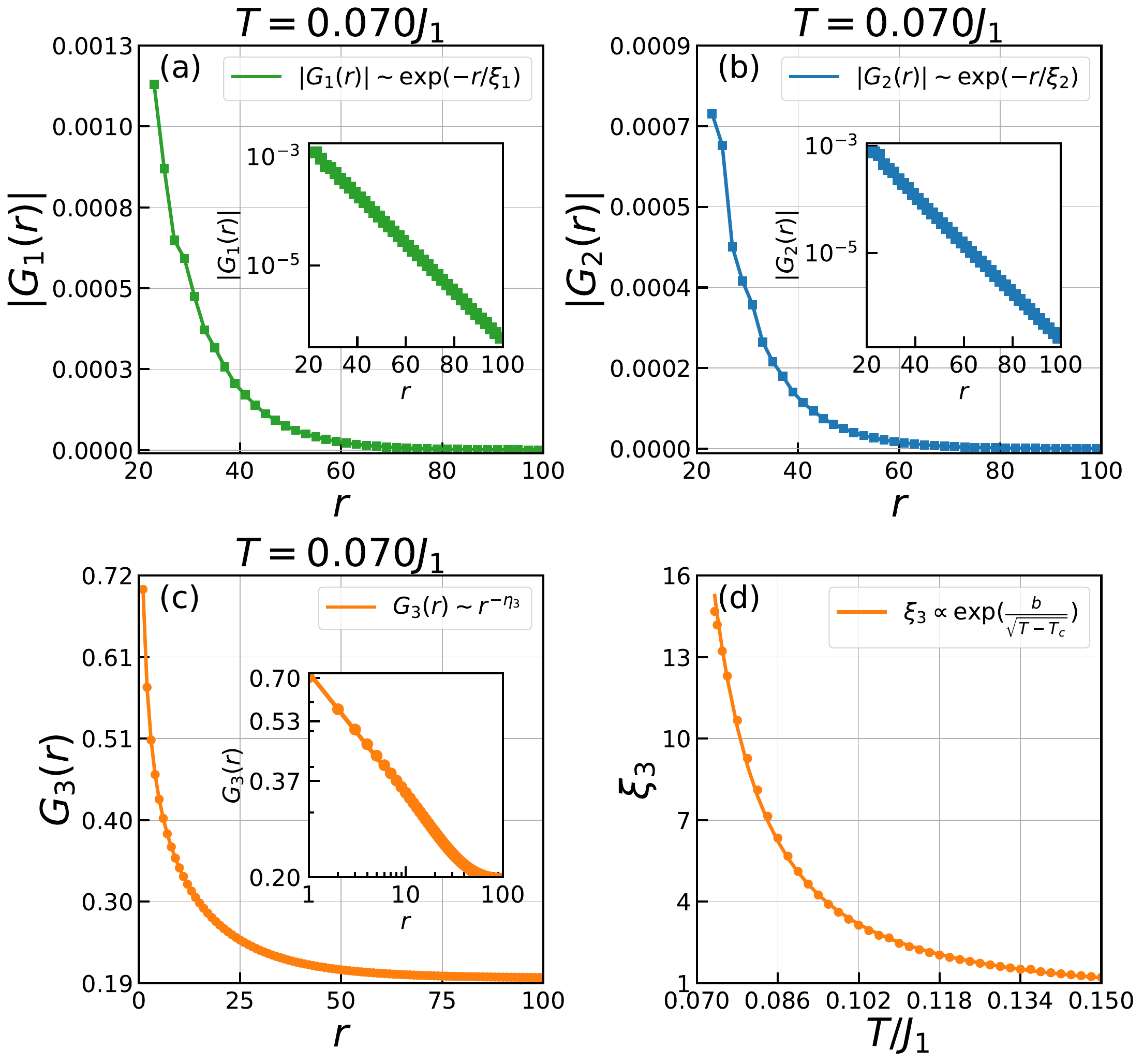}
\caption{(a)-(c) The correlation function in the charge 6e coherent phase, at $J_2=0.0025J_1,T=0.070J_1$. (a) The correlation function of the integer vortices. (b) The correlation
function of the $1/2$ vortices. (c) The correlation function of the $1/3$
vortices. (d) The extracted correlation length of $\frac{1}{3}$ vortices.}
\label{fig: correlation function}
\end{figure}

The most intriguing physics occurs at $J_2<0.005J_1$, where a vestigial $\frac{1}{3}$ vortex-antivortex paired phase emerges. The intermediate phase at $T_{c1}<T<T_{c2}$, can be characterized by the following correlation function
\begin{eqnarray}
&&G_n(r)=\langle e^{in(\theta_i-\theta_{i+r})}\rangle=|G_n(r)|e^{i\phi_n(r)},
\end{eqnarray}
where $G_n(r)$ describes the correlation behavior of $1/n$ fractional
vortices when $n>1$. In Fig.\ref{fig: correlation function}(a)-(b), the
correlation functions for both integer vortices and $1/2$ vortices are
calculated at $T=0.070J_1$, and their amplitudes exhibit the exponential
decay. In contrast,, the correlation function for $\frac{1}{3}$ fractional vortices exhibits QLRO:
\begin{equation}
G_3(r)=\langle e^{3i(\theta_i-\theta_{i+r})}\rangle\sim r^{-\eta_3},
\end{equation}
as depicted in Fig.\ref{fig: correlation function}(c). It is noteworthy that $G_3(r)$
at large distances shows a tail due to the finite bond dimension. Moreover,
above $T_{c2}$, a correlation length can be determined with the form of $
G_3(r)\sim \exp(-r/\xi_3)$. As shown in Fig.\ref{fig: correlation function}
(d), we have
\begin{equation}
\xi_3(T)\sim\exp (\frac{b}{\sqrt{T-T_c}}),\quad T\rightarrow T_{c}^{+},
\end{equation}
suggesting a BKT phase transition at $T_{c2}$ driven by
the $\frac{1}{3}$ vortex unbinding. These findings confirm the existence of a vestigial charge-$6e$ coherent phase, where the $6e$ correlation exhibits QLRO, in the absence of any $2e$ correlation.

The first-order transition between the charge-$2e$ PDW phase and the charge-$6e$ coherent phase can be understood through the properties of the order parameter degeneracy space~\cite{Korshunov_1986}. The degeneracy space of the order parameter for the charge-$2e$ PDW phase is $U(1) \times Z_2$, whereas for the charge-$6e$ coherent phase, the symmetry is enlarged to include a discrete $Z_3$ subgroup, where $Z_3$ corresponds to rotations by multiples of $2\pi/3$. As a result, the symmetry restored by the domain walls is not just $Z_2$ but $Z_2 \times Z_3$, corresponding to arbitrary permutations of $\theta_A$, $\theta_B$, and $\theta_C$. This symmetry structure leads to a phase transition in the universality class of the six-state Potts model, which is characterized by a first-order transition.

\section{Discussion and Conclusion}

We have conducted a detailed investigation into the defect-driven melting of commensurate PDW states using a combination of GL theory with tensor network techniques. Within the GL framework, we identify the $\psi_{\text{kagome}}$ state, consistent with the $2a_0 \times 2a_0$ $3Q$ spatial superconducting modulations and TRSB observed in experiments~\cite{Deng_2024, Tian_2023}. Our analysis of the superconducting order parameter reveals two key topological defects—domain walls and $1/3$ vortices—that play a crucial role in phase fluctuations at finite temperatures. We find that the essential physics governing these topological excitations can be captured by an effective antiferromagnetic $J_1-J_2$ XY model on the emergent kagome lattice. To solve this effective model, we developed a tensor network method and performed a detailed study in the regime $J_2 \ll J_1$, which is challenging for conventional Monte Carlo methods.

It has been found that for $J_2/J_1 < 0.005$, as the temperature increases, the system undergoes a first-order transition from the $\psi_{\text{kagome}}$ state to an intermediate $1/3$ vortex-paired phase, driven by the proliferation of domain walls. As the temperature increases further, the system transitions to the normal state via a BKT transition, due to the dissociation of fractional vortices. In contrast, for $J_2/J_1 > 0.005$, only a single first-order phase transition occurs, with no intermediate fractional vortex-paired phase.

In the context of superconducting (SC) phase fluctuations, the quasi-long-range order (QLRO) of $e^{i\theta}$ corresponds to the global phase coherence of $2e$ Cooper pairs, while the QLRO of $e^{i3\theta}$ reflects the phase coherence of the charge $6e$ order parameter. The intermediate fractional vortex-paired phase can then be understood as a charge $6e$ coherent phase, in which the phase coherence is established between the charge $6e$ order parameter, in the absence of charge $2e$ phase coherence.

Our work provides a plausible explanation for the experimental results observed in $\rm{CsV_3Sb_5}$~\cite{Ge_2023}. In the experiments, the resistance data reveal a broad SC transition region, indicating strong SC fluctuations. Magnetoresistance oscillations were reported in five $\rm{CsV_3Sb_5}$ ring device samples (s1 $\sim$ s5) of varying sizes. In the samples with smaller sizes (s1, s3), flux quantization signals were observed at units of $h/2e$, $h/4e$, and $h/6e$. In contrast, in the larger samples (s2, s4, s5), only flux quantization at $h/2e$ and $h/6e$ was detected.

The observed transition from $h/2e$ to $h/6e$ oscillations in the ring devices can be understood as a result of thermal phase transitions between the pair density wave (PDW) phase and the charge-$6e$ coherent phase. Notably, our theoretical framework predicts a first-order transition between these two phases, which could be further explored experimentally. However, an unresolved issue remains regarding the $\frac{h}{4e}$ oscillation detected in the s1 samples~\cite{Ge_2023}. Our current model, which focuses on phase fluctuations within the $\psi_{\text{kagome}}$ state, does not account for this anomaly. The charge-$4e$ signal is notably weaker than the $2e$ and $6e$ signals, and is observed only in the smaller samples, suggesting that it may arise from alternative mechanisms that require further investigation.

The emergence of $\frac{1}{3}$ vortices in these states is a natural consequence of the interplay between time-reversal symmetry and $U(1)$ symmetry in the PDW state. This behavior is different from the incommensurate case, where fractional vortices arise from the $U(1) \times U(1) \times U(1)$ degeneracy~\cite{Agterberg_2011}. Notably, the combination of continuous and discrete degeneracies serves as a general mechanism for the formation of fractional vortices in frustrated XY models~\cite{Korshunov_2006}. For commensurate PDW states, similar degeneracies emerge, where discrete symmetry breaking can result from either time-reversal symmetry, translational symmetry breaking or rotational symmetry breaking. The presence of fractional vortices, along with the possibility of an intermediate fractional vortex-pair phase during the partial melting of PDW states, presents a compelling direction for further exploration.

\textbf{Acknowledgments.}
-The authors are very grateful to Ziqiang Wang stimulating discussions. The research is supported by the National Key
Research and Development Program of China (Grant No. 2023YFA1406400).

\appendix
\counterwithin{figure}{section}

\section{Commensurate PDW Order}\label{sec:phi_ruby}
By imposing the translational symmetry of $2a_0$ for the commensurate PDW, along with the symmetries of the space group $P6/mmm$, $U(1)$, and time-reversal symmetry, we obtain the Ginzburg-Landau (GL) free energy density up to fourth order:
\begin{equation}
F(\{\Delta_{\bm{Q}}\}) = F_I(\{\Delta_{\bm{Q}}\}) + F_c^{(2)}(\{\Delta_{\bm{Q}}\}) + F_c^{(4)}(\{\Delta_{\bm{Q}}\}),
\end{equation}
where $F_I$ represents the incommensurate part, and $F_c^{(2)}$ and $F_c^{(4)}$ correspond to the second- and fourth-order terms of the commensurate part, respectively.

For $\beta_5 > 0$, the ground state with TRSB is energetically favored. Two possible candidates for the ground state are the $\psi_{\text{kagome}}$ state
\begin{equation}
\psi_{\text{kagome}}^{\pm}(r)=\Delta e^{i\theta} \sum_{j=1,2,3} e^{\pm i
(j-2) \frac{\pi}{3}}\cos [\bm{Q}_j\cdot \bm{r}+\phi_j],
\label{eqn: kagome pdw}
\end{equation}
and the $\psi_{\text{Ruby}}$ state
\begin{equation}
\psi_{\text{Ruby}}^{\pm}(r)=\Delta e^{i\theta} \sum_{j=1,2,3} e^{\pm i
(j-2) \frac{\pi}{3}}\sin [\bm{Q}_j\cdot \bm{r}+\phi_j].
\label{eqn: Ruby pdw}
\end{equation}
Since the $\psi_{\text{kagome}}$ state has been discussed extensively in the main text, here we demonstrate that the $\psi_{\text{Ruby}}$ state exhibits similar properties.

\begin{figure}[t]
\centering
\includegraphics[width=\linewidth]{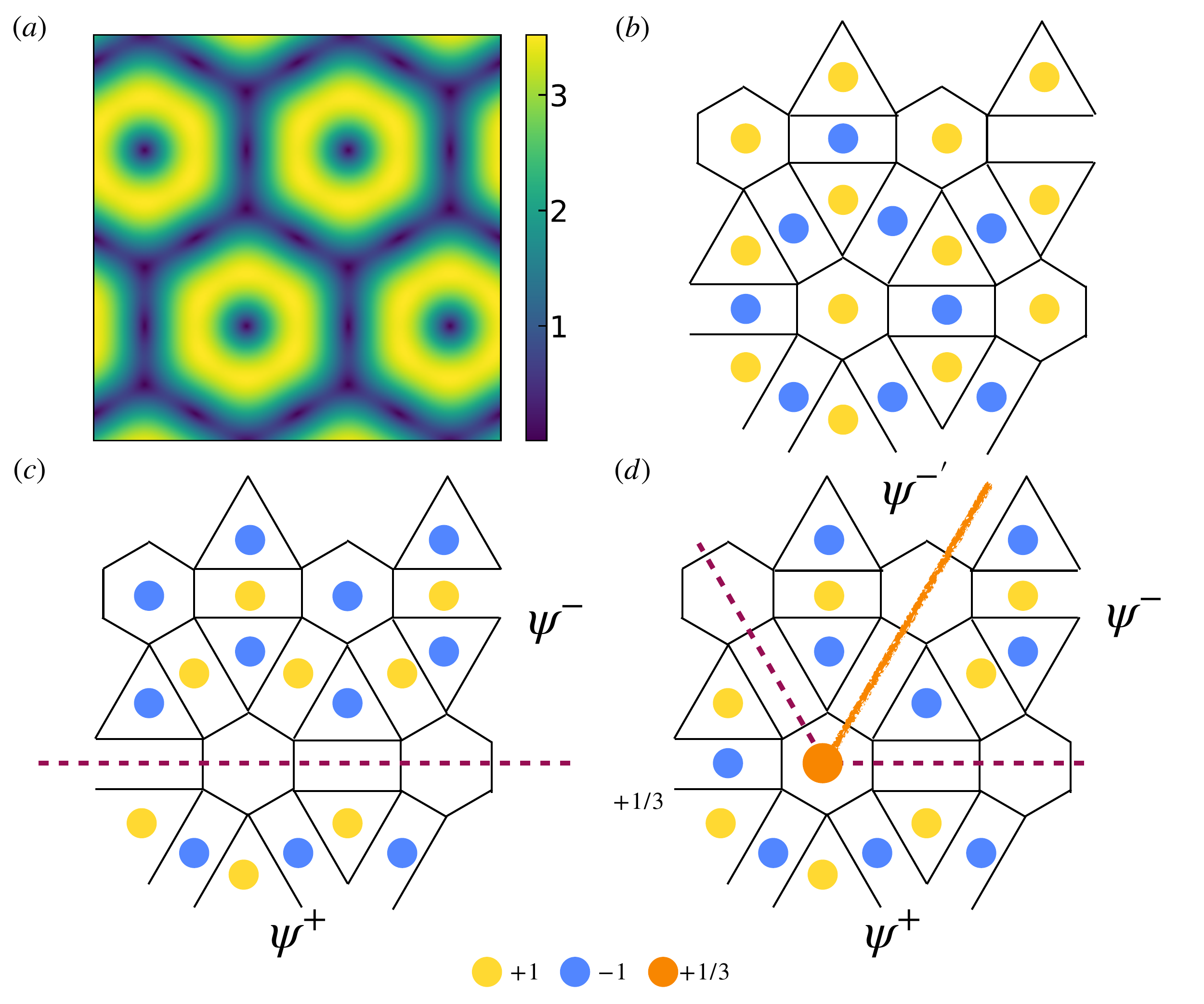}
\caption{(a) The amplitude distribution of $\protect\psi_{\text{Ruby}}$.
(b) The vorticity distribution in $\protect\psi_{\text{Ruby}}$. The phase distribution of $\protect\psi_{\text{Ruby}}$ on the emergent Ruby lattice. (c) Order parameter with an infinite straight domain wall, depicted by the purple dashed line. (d) The $+\frac{1}{3}$ vortex excitation at the kink of the domain wall.}
\label{fig:Ruby_ground_state}
\end{figure}

In Fig.~\ref{fig:Ruby_ground_state}(a), we illustrate the amplitude distribution of the superconducting order parameter $\psi_{\text{Ruby}}^{+}(r)$. In this case, the maxima of $\psi_{\text{Ruby}}^{+}(r)$ form a Ruby lattice. As shown in Fig.~\ref{fig:Ruby_ground_state}(b), each triangular and hexagonal plaquette hosts a vortex with vorticity $\pm 1$, while rectangular plaquettes contain vortices with vorticity $\mp 1$.

The topological defects in the $\psi_{\text{Ruby}}$ state can be constructed in a similar manner as the kagome case. As illustrated in Fig.~\ref{fig:Ruby_ground_state}(c), an infinite domain wall can be formed along the centers of the hexagonal and rectangular plaquettes
\begin{equation}
    \bm{Q}_j \cdot \bm{r} = \left(n + \frac{1}{2}\right)\pi, \quad n \in \mathbb{Z},
\end{equation}

If the state below the domain wall is given by
\begin{equation}
    \psi_{\text{Ruby}}^{+}(r) = \Delta \sum_{j=1,2,3} e^{+i(j-2)\frac{\pi}{3}} \sin(\bm{Q}_j \cdot \bm{r} + \phi_j), \label{eqn: domain wall excitation 4}
\end{equation}
then the energy of the system is minimized when the order parameter on the opposite side of the wall is
\begin{equation}
    \psi_{\text{Ruby}}^{-}(r) = \Delta e^{i\frac{2\pi}{3}} \sum_{j=1,2,3} e^{-i(j-2)\frac{\pi}{3}} \sin(\bm{Q}_j \cdot \bm{r} + \phi_j).\label{eqn: domain wall excitation 5}
\end{equation}
Here, across the domain wall, the vorticity in each plaquette is reversed.

Moreover, topological defects corresponding to $1/3$ vortices can be found at the intersection points of two domain walls. Consider two half-infinite domain walls defined by
\begin{equation}
    \bm{Q}_2 \cdot \bm{r} = \left(n + \frac{1}{2}\right)\pi, \quad \bm{Q}_3 \cdot \bm{r} = \left(m + \frac{1}{2}\right)\pi,
\end{equation}
as illustrated in Fig.~\ref{fig:Ruby_ground_state}(d). These domain walls intersect at the center of a hexagon, creating a $\frac{2\pi}{3}$ kink. After buckling, the state above the domain wall is given by
\begin{equation}
   \psi_{\text{Ruby}}^{-'}(r) = \Delta \sum_{j=1,2,3} e^{-i(j-2)\frac{\pi}{3}} \sin(\bm{Q}_j \cdot \bm{r} + \phi_j), \label{eqn: domain wall excitation 6}
\end{equation}
resulting in a global $\frac{2\pi}{3}$ phase mismatch along the orange line.

\section{Tensor Network method}
\label{sec: tensor}
\subsection{Tensor network representation of the partition function}
The Hamiltonian of the $J_1-J_2$ antiferromagnetic XY model on a kagome lattice is defined as
\begin{equation}
    H=J_1\sum_{\langle i,j\rangle}\cos(\theta_i-\theta_j)+J_2\sum_{\langle\langle i,j\rangle\rangle}\cos(\theta_i-\theta_j),
    \label{eqn:hamiltonian SM}
\end{equation}
where $J_1$ and $J_2$ represent the coupling constants of nearest neighbor and next nearest neighbor interactions, respectively. At a given temperature $T$, the partition function is given by:
\begin{align}
\begin{split}
Z&=\prod_{i}\int{\frac{d\theta_i}{2\pi}}\exp(-\beta H({\{\theta_i\}}))\\
&=\prod_{i}\int{\frac{d\theta_i}{2\pi}}\prod_{\langle i,j\rangle }e^{-k_1\cos(\theta_i-\theta_j)}\prod_{\langle \langle i,j\rangle\rangle}e^{-k_2\cos(\theta_i-\theta_j)},
\end{split}
\end{align}
where $k_1=J_1/T$ and $k_2=J_2/T$.

\begin{figure}[t]
\centering
\includegraphics[width=\linewidth]{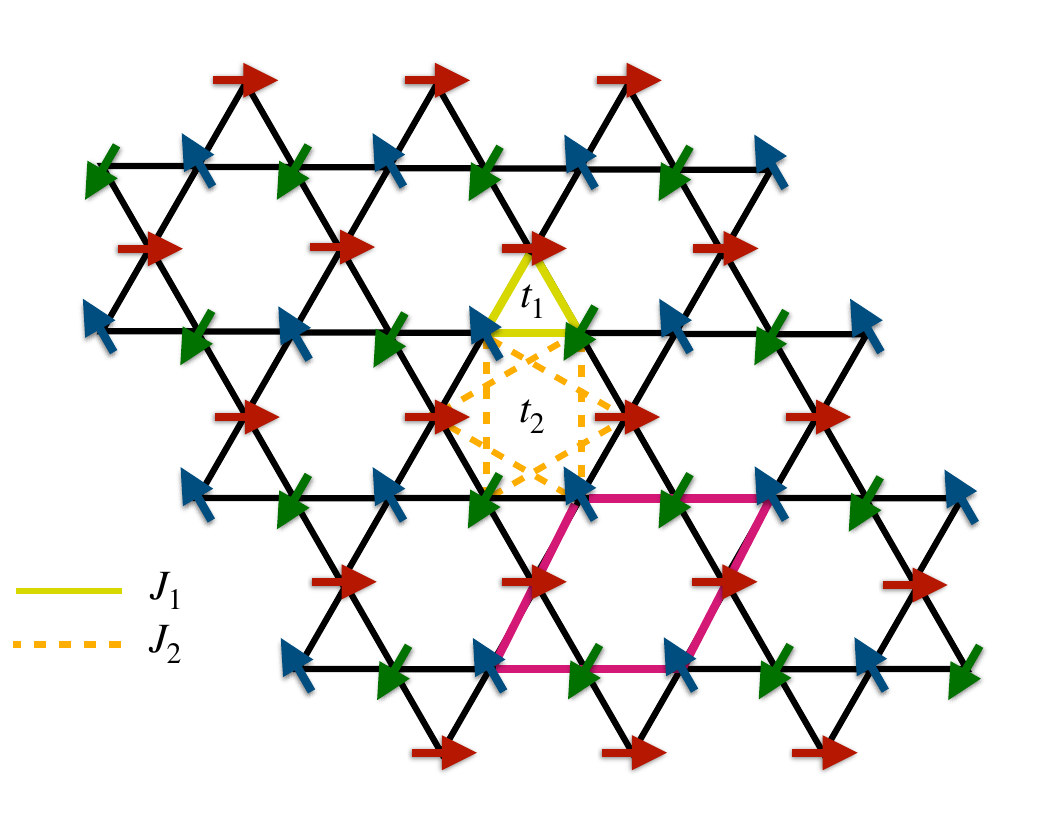}
\caption{Classical $J_1-J_2$ antiferromagnetic XY model on a kagome lattice and its ground state. The unit cell of the kagome lattice is shown as the pink parallelogram. The NN triangle $t_1$ and NNN triangles are shown by yellow and orange lines, respectively. }
\label{fig:ground_state}
\end{figure}

In the TN formalism, the first step in the calculation is to represent the partition function as an infinite TN. For the non-frustrated XY model, the TN representation can be directly constructed via Fourier transformation on each two-body term, known as the standard construction \cite{Jos_1977,Vanderstraeten_2019,Yu_2014}:
\begin{equation}
    e^{-k_i\cos(\theta_i-\theta_j)}=\sum_{n_{ij}=-\infty}^{\infty}I_{n_{ij}}(-k_i)e^{in (\theta_i-\theta_j)},
    \label{eqn:standard construction}
\end{equation}
where $I_n$ represents the modified Bessel function of the first kind. However, for the frustrated model, the standard TN representation fails to converge. Since the $J_1-J_2$ AFXY kagome lattice model is a frustrated system, constructing its TN representation requires careful consideration \cite{Vanhecke_2021,Song_2023_2}. In our previous work \cite{Song_2023_2}, we established a comprehensive framework for addressing frustrated models with continuous degrees of freedom. This framework can be seamlessly applied to the present case, providing a natural and effective approach.

\begin{figure*}[t]
\centering
\includegraphics[width=\linewidth]{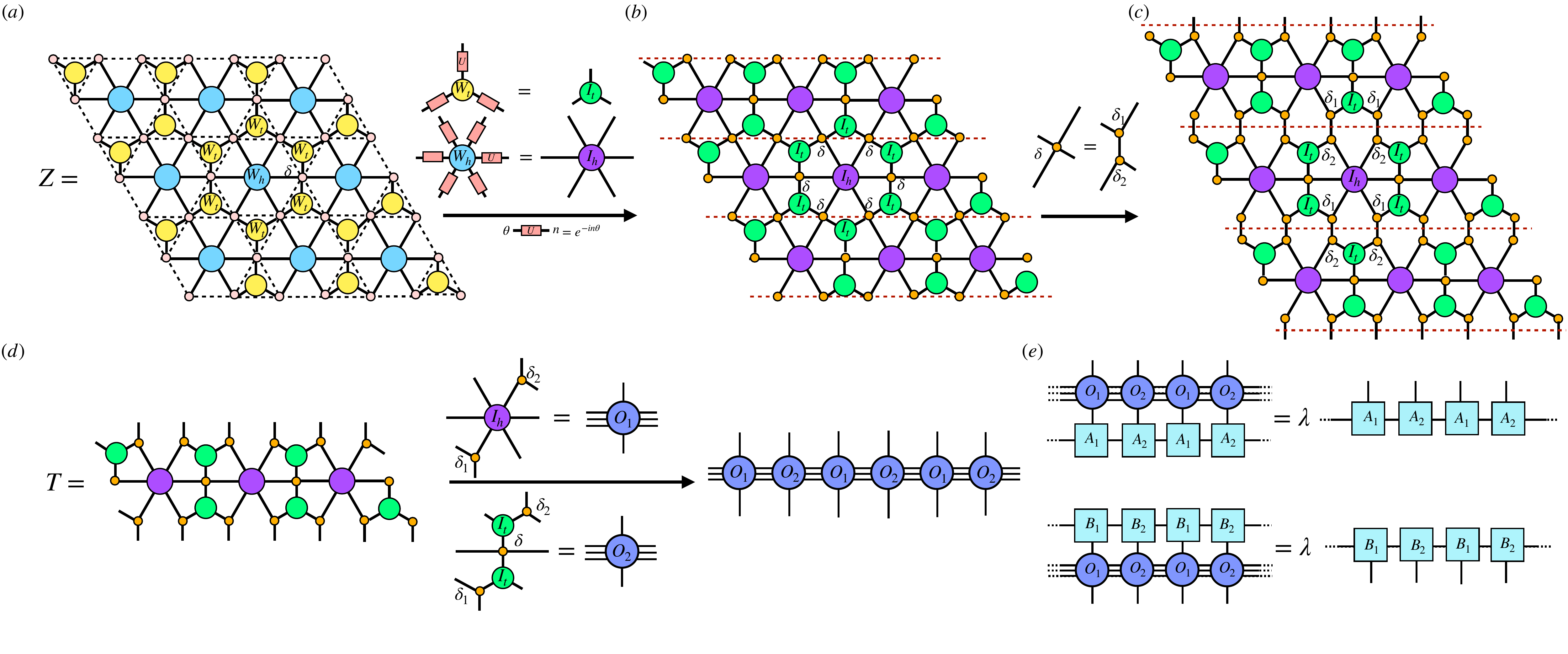}
\caption{TN representation of the $J_1-J_2$ AFXY model.
The relationship between local tensors is depicted on the arrows connecting different tensor network diagrams.
(a) TN with continuous indices. The dashed lines represent the bonds of the original kagome lattice, while solid lines depict the legs of tensors.
(b) TN with discrete indices constructed by Fourier transformation on each plaquette and integration of spin variables.
(c) The vertical splitting of the $\delta$ tensor, with the split bonds shown by red dashed lines.
(d) Construction of the transfer matrix and uniform local tensors.
(e) Up and down eigenequations for the fixed-point uniform matrix product state (MPS) $|\psi(A)\rangle$ and $\langle \psi(B)|$ of the transfer matrix $T$.}
\label{fig:TN_rep}
\end{figure*}

To achieve convergence, the local tensor of the TN representation should satisfy the ground state local rule \cite{Vanhecke_2021,Song_2023_2}. As shown in Fig.\ref{fig:ground_state}, The three spins on both the nearest neighbor triangle ($t_1$) and next nearest neighbor triangle ($t_2$) forming a relative $\frac{2\pi}{3}$ phase difference with each other. Hence, we reformulate the Hamiltonian into local terms acting on clusters of spins
\begin{equation}
    H=\sum_{t_1} H_{t_1}+\sum_{t_2} H_{t_2},
\end{equation}
where $H_{t_1}$ contains all the interaction terms in the nearest neighbor triangle, and $H_{t_2}$ contains all the interaction terms in the next nearest neighbor triangles. The partition function can then be reformulated as:
\begin{equation}
    Z=\prod_{i}\int{\frac{d\theta_i}{2\pi}}\prod_{t_1}W_{t_1}(\{\theta_{t_1}\})\prod_{t_2}W_{t_2}(\{\theta_{t_2}\}),
\end{equation}
where
\begin{eqnarray}
    &&W_{t_1}(\{\theta_{t_1}\})=W_{t_1}(\theta_1,\theta_2,\theta_3)=
    \exp[-k_1(\cos(\theta_1-\theta_2)\notag \\
    &&+\cos(\theta_2-\theta_3)+\cos(\theta_3-\theta_1))],
\end{eqnarray}
and
\begin{eqnarray}
    &&W_{t_2}(\{\theta_{t_2}\})=W_{t2}(\theta_1,\theta_2,\theta_3)=
    \exp[-k_2(\cos(\theta_1-\theta_2)\notag \\
    &&+\cos(\theta_2-\theta_3)+\cos(\theta_3-\theta_1))],
\end{eqnarray}
represent the Boltzmann weights on nearest neighbor triangles and next nearest neighbor triangles.
As each hexagonal plaquette contains two next-nearest neighbor triangles, we can define
\begin{equation}
    W_h(\theta_1,\theta_2,\theta_3,\theta_4,\theta_5,\theta_6)=W_{t_2}(\theta_1,\theta_3,\theta_5)W_{t_2}(\theta_2,\theta_4,\theta_6).
\end{equation}
The partition function can then be written as
\begin{equation}
    Z=\prod_{i}\int{\frac{d\theta_i}{2\pi}}\prod_{t_1}W_{t_1}(\{\theta_{t_1}\})\prod_{h}W_h(\{\theta_h\}).
\end{equation}
This partition function can be translated into a TN with continuous $U(1)$ indices. As shown in Fig. ~\ref{fig:TN_rep}(a), $W_t$ can be viewed as a three-leg tensor with continuous indices located at the center of triangle plaquettes and $W_h$ is a six-leg tensor located at the center of hexagonal plaquettes. These local tensors, sitting on the dual lattice, are connected by the $\delta$ tensor which is located at the original lattice, representing that each physical spin is shared by four neighboring clusters of Boltzmann weight.
The $\delta$ tensor can be expressed as
\begin{equation}
    \delta({\theta_1,\theta_2,\theta_3,\theta_4})=\int\frac{d\theta_0}{2\pi}\delta(\theta_1-\theta_0)\delta(\theta_2-\theta_0)\delta(\theta_3-\theta_0)\delta(\theta_4-\theta_0).
\end{equation}

In order to transform the local tensors onto a discrete basis, we make a Fourier transformation on Boltzmann weight $W_{t}$
\begin{eqnarray}
&&I_{t_1}(n_1,n_2,n_3)=\notag\\
&&\prod_{i=1}^{3}\int\frac{d\theta_i}{2\pi}W_{t_1}(\theta_1,\theta_2,\theta_3)U(\theta_1,n_1)U(\theta_2,n_2)U(\theta_3,n_3),\notag\\
\label{eqn: FT for triangular plaquette}
\end{eqnarray}
and $W_{h}$
\begin{eqnarray}
   &&I_h(n_1,n_2,n_3,n_4,n_5,n_6) \notag\\
   &&=(\prod_{i=1}^{6}\int\frac{d\theta_i}{2\pi})W_h(\theta_1,\theta_2,\theta_3,\theta_4,\theta_5,\theta_6)
   \prod_{i=1}^{6}U(\theta_i,n_i) \notag \\
   &&=(\prod_{i=1,3,5}\int\frac{d\theta_i}{2\pi})W_{t_2}(\theta_1,\theta_3,\theta_5)\prod_{i=1,3,5}U(\theta_i,n_i)\times \notag\\
   &&(\prod_{i=2,4,6}\int\frac{d\theta_i}{2\pi})W_{t_2}(\theta_2,\theta_4,\theta_6)\prod_{i=2,4,6}U(\theta_i,n_i) \notag\\
   &&=I_{t_2}(n_1,n_3,n_5)I_{t_2}(n_2,n_4,n_6),
\end{eqnarray}
where
\begin{equation}
U(\theta,n)=e^{-in\theta},
\end{equation}
is the basis of Fourier transformation.
The original phase variables $\{\theta\}$ can then be integrate out on each site
and the partition function is transformed into the TN representation shown in Fig.~\ref{fig:TN_rep}(b)
\begin{eqnarray}
    &&Z=\lim_{N\rightarrow\infty}\prod_{t_1}(\prod_{i=1}^{3}\sum_{n_{t_1}^i=-N}^{N}I_{t_1}(\{n_{t_1}^i\}))\notag \times\\
    &&\prod_{h}(\prod_{i=1}^{6}\sum_{n_h^i=-N}^{N}I_h(\{\theta^i_h\}))\prod_{s}\delta(n_s^1,n_s^2,n_s^3,n_s^4),\notag\\
\end{eqnarray}
where $\delta$ tensor is a Kronecker delta function defined by
\begin{equation}
\delta(n_1,n_2,n_3,n_4)=\int\frac{d\theta}{2\pi}\prod_{i=1}^{4}U(n_i,\theta)=\delta_{n_1+n_2+n_3+n_4,0}. \label{eqn: delta_tensor}
\end{equation}

To construct the linear transfer matrix of the TN, we decompose the $\delta$ tensor vertically as is shown in Fig.~\ref{fig:TN_rep}(c):
\begin{equation}
    \delta_{n_1+n_2+n_3+n_4,0}=\sum_{n_0}\delta_{n_1+n_2-n_0,0}^{1}\delta_{n_3+n_4+n_0,0}^{2}.
    \label{eqn: vertical_split U(1) spin}
\end{equation}
The transfer matrix $T$ can be built by grouping the $\{\delta^1,\delta^2,\delta,I_h,I_t\}$ tensors and contract them as is shown in Fig.~\ref{fig:TN_rep}(d).

We can further split the transfer matrix horizontally to obtain the local uniform tensors
$O_1$ and $O_2$ whose interior structures are shown in Fig.~\ref{fig:TN_rep}(d). The resulting TN representation $N$ is given by
\begin{equation}
    Z=\text{tTr}(\dots TTT\dots)=\text{tTr}(\prod_{s\in N}O_1(s)O_2(s)),
\end{equation}
where $s$ denotes the unit cell in the network.

\subsection{Finite bond dimension truncation of the local tensors}
\begin{figure}[t]
\centering
\includegraphics[width=0.8\linewidth]{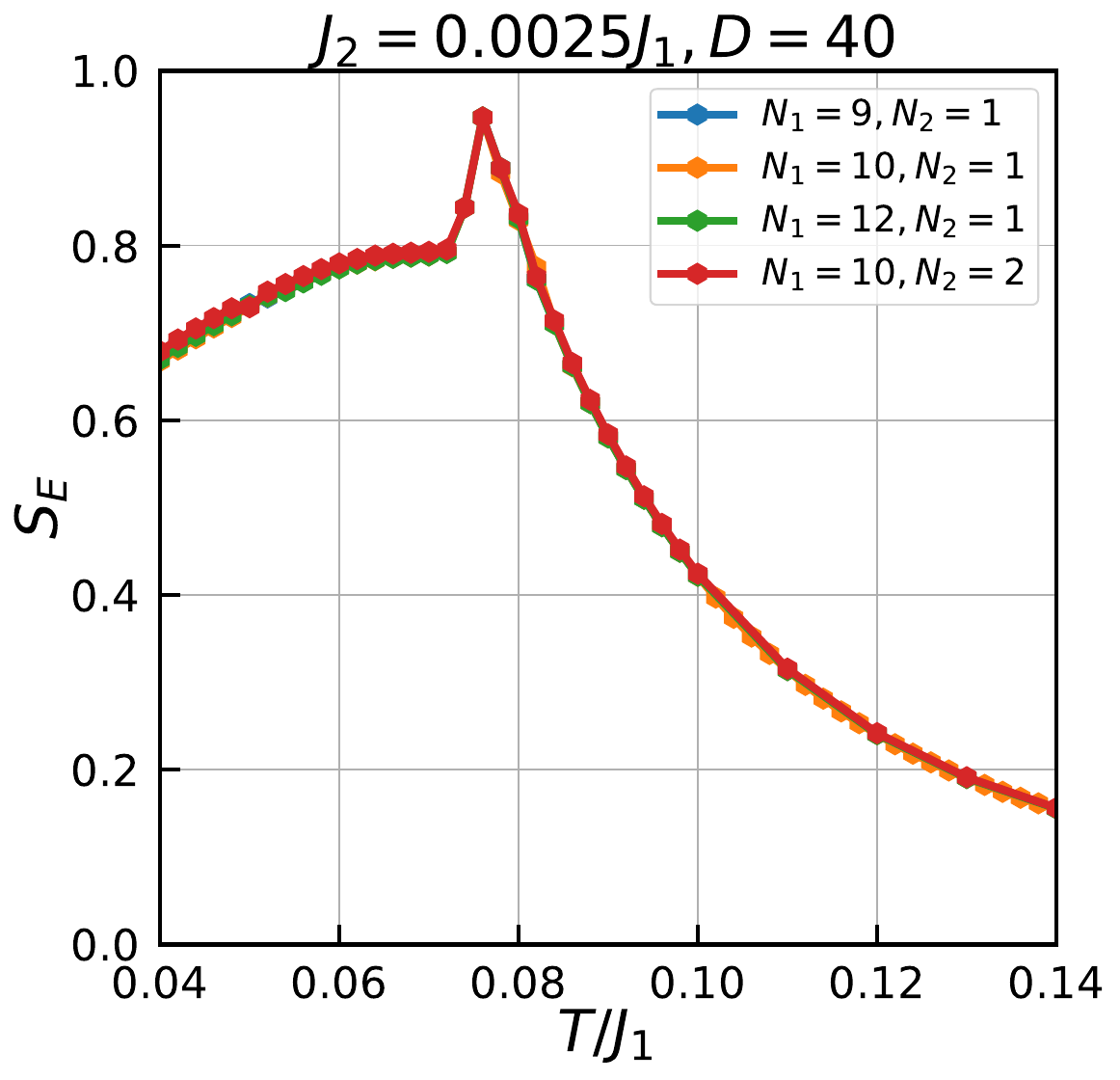}
\caption{The generalized entanglement entropy as a function of temperature at $J_2=0.0025J_1$, with varying truncation dimensions for local tensors and fixed MPS bond dimension $D=40$.}
\label{fig:EE_n1_n2}
\end{figure}

While the tensor network with discrete indices is obtained, for practical calculations, it is necessary to truncate the indices $n_i$ to a finite range: $-N_{\text{max}} \leq n_i \leq N_{\text{max}}$. To show the effectiveness of truncation, we derive the asymptotic form of local tensors
\begin{widetext}
\begin{eqnarray}
     &&I_t(n_1,n_2,n_3)
    =\prod_{i=1}^{3}\int\frac{d\theta_i}{2\pi}\exp(-K\cos(\theta_1-\theta_2)-K\cos(\theta_2-\theta_3)-K\cos(\theta_3-\theta_1))\times e^{in_1\theta_1}e^{in_2\theta_2}e^{in_3\theta_3} \notag \\
    &&\sim\int\frac{dx}{2\pi}\int\frac{dy}{2\pi}\int\frac{d\theta_3}{2\pi}\exp(-K\cos(x)-K\cos(y)-K\cos(x+y))e^{in_1x}e^{i(n_1+n_2)y}e^{i(n_1+n_2+n_3)\theta_3}\notag\\
    &&=\int\frac{dx}{2\pi}\int\frac{dy}{2\pi}\exp(-K\cos(x)-K\cos(y)-K\cos(x+y))e^{in_1x}e^{i(n_1+n_2)y}\delta(n_1+n_2+n_3).
\end{eqnarray}
Here we make a variable substitution: $\theta_1-\theta_2=x,\theta_2-\theta_3=y$. As the local hamiltonian on triangle have two minimum, $x=y=\pm\frac{2\pi}{3}$, we make saddle point approximation around $x=y=\pm\frac{2\pi}{3}\sigma_t,\sigma_t=\pm 1$:
\begin{eqnarray}
    I_t(n_1,n_2,n_3)\sim && \sum_{\sigma_t=\pm 1}\int\frac{dx}{2\pi}\int\frac{dy}{2\pi}\exp(\frac{1}{4}(x-\frac{2\pi}{3}\sigma_t)^2+\frac{1}{4}(y-\frac{2\pi}{3}\sigma_t)^2+\frac{1}{4}(x-\frac{2\pi}{3}\sigma_t+y-\frac{2\pi}{3}\sigma_t)^2) \notag\\ &&\times e^{in_1x}e^{i(n_1+n_2)y}\delta(n_1+n_2+n_3).
\end{eqnarray}
By using Gaussian integral formula, we obtain:
\begin{eqnarray}
    &&I_t(n_1,n_2,n_3)\sim\sum_{\sigma_t=\pm 1}\exp(i\frac{2\pi}{3}\sigma_t(2n_1+n_2)-\frac{1}{3K}n_1^2-\frac{1}{3K}(n_1+n_2)^2-\frac{1}{3K}n_2)^2\delta(n_1+n_2+n_3).\notag\\
    &&=\sum_{\sigma_t=\pm 1}\exp(i\frac{2\pi}{3}\sigma_t(n_1+n_2+n_3)-\frac{1}{3K}(n_1^2+n_2^2+n_3)^2\delta(n_1+n_2+n_3).
\end{eqnarray}
Since the tensor elements decay exponentially with tensor indices $n_i$, we can safely truncate the local tensors with finite bond dimensions. Besides, since $J_1\gg J_2$, a larger truncation bond dimension is needed for $I_{t_1}$ and a smaller one for $I_{t_2}$. As shown in Fig.(\ref{fig:EE_n1_n2}), the generalized entanglement entropy remains nearly unchanged under different bond dimensions of the local tensors from $N_1=9\sim 12, N_2=1\sim2$. In practice, we choose $N_{\text{max}}=10$ for $I_{t_1}$ and $N_{\text{max}}=1$ for $I_{t_2}$ for the calculation of physical quantities. Note that, while the standard construction becomes a poor approximation and eventually fails at low temperatures after the finite bond dimension cutoff \cite{Song_2023}, Eq.(\ref{eqn: FT for triangular plaquette}) remains effective.

To compare with the standard construction, we perform a standard expansion for $I_t$ tensor:
\begin{eqnarray}
       &&I_{t_i}(n_1,n_2,n_3)=
\prod_{i=1}^{3}\int\frac{d\theta_i}{2\pi}\exp[-k_i\cos(\theta_1-\theta_2)]\exp[-k_i\cos(\theta_2-\theta_3)]\exp[-k_i\cos(\theta_3-\theta_1)]e^{in_1\theta_1}e^{in_2\theta_2}e^{in_3\theta_3}\notag\\
&&=\prod_{i=1}^{3}\int\frac{d\theta_i}{2\pi}
(\sum_{n_{12}}I_{n_{12}}(-k_i)e^{in_{12}(\theta_1-\theta_2)})
\times(\sum_{n_{23}}I_{n_{23}}(-k_i)e^{in_{23}(\theta_2-\theta_3)})\times(\sum_{n_{31}}I_{n_{31}}(-k_i)e^{in_{31}(\theta_3-\theta_1)})\prod_{i=1}^{3}e^{in_i\theta_i}\notag\\
&&=\sum_{n_{12}}\sum_{n_{23}}\sum_{n_{31}}I_{n_{12}}(-k_i)I_{n_{23}}(-k_i)I_{n_{31}}(-k_i)\times\delta(n_{12}-n_{31}+n_1)\delta(n_{23}-n_{12}+n_2)\delta(n_{31}-n_{23}+n_3)\notag\\
&&=\sum_{n_{12}}I_{n_{12}}(-k_i)I_{n_{12}-n_2}(-k_i)I_{n_{12}+n_{1}}(-k_i)\delta(n_1+n_2+n_3).
\label{eqn: compared with standard construction}
\end{eqnarray}
\end{widetext}

Compared with the standard constructions, when making cut-off on $n_i$, Eq.(\ref{eqn: compared with standard construction}) includes an additional infinite summation over $n_{ij}$ on each triangle, making it effective at low temperatures.

\subsection{Solving the leading eigenvalue of the transfer matrix}
After the TN representation is obtained, the network can be contracted by solving the leading eigenequations for the transfer matrix $T$ given by
\begin{equation}
T(k_1,k_2)|\psi_{k_1,k_2}^d\rangle =\Lambda|\psi_{k_1,k_2}^d\rangle,
\label{eqn:eigen_equation_A}
\end{equation}
\begin{equation}
\langle\psi_{k_1,k_2}^u|T(k_1,k_2) =\Lambda\langle\psi_{k_1,k_2}^u|,
\label{eqn:eigen_equation_B}
\end{equation}
where $|\psi^d_{k_1,k_2}\rangle$ is the leading down eigenvector of the transfer matrix and $\langle\psi^u_{k_1,k_2}|$ is the leading up eigenvector. Since the transfer matrix $T$ is non-hermitian, both the up and down leading eigenvectors need to be solved.

These leading eigenvectors can be accurately approximated using a uniform matrix product state \cite{Laurens_2019}. As the transfer matrix has a $1\times 2$ unit cell, we utilize a multi-site uniform matrix product state described by two tensors $A_1$ and $A_2$:
\begin{eqnarray}\label{eqn: up eigenequation}
    |\psi(A_1,A_2)\rangle=
    \sum_{\{n_i\}}&Tr(\cdots A_1^{n_1}A_2^{n_2}A_1^{n_3}A_2^{n_4}\cdots) \notag\\
    &|\cdots n_1,n_2,n_3,n_4,\cdots\rangle,
\end{eqnarray}
where $A_1$ and $A_2$ are three-leg tensors with a physical bond dimension $d$ determined by the transfer matrix and a virtual bond dimension $D$ controlling the accuracy of the MPS approximation. The up eigenvector $\langle\psi^u|$ is represented by another MPS $\langle\psi(B_1,B_2)|$:
\begin{eqnarray}\label{eqn: down eigenequation}
    \langle\psi(B_1,B_2)|=
    \sum_{\{n_i\}}&Tr(\cdots B_1^{n_1}B_2^{n_2}B_1^{n_3}B_2^{n_4}\cdots) \notag\\
    &\langle\cdots n_1,n_2,n_3,n_4,\cdots|.
\end{eqnarray}

\subsection{Spatial symmetry of transfer matrix}
\begin{figure}[t]
\centering
\includegraphics[width=\linewidth]{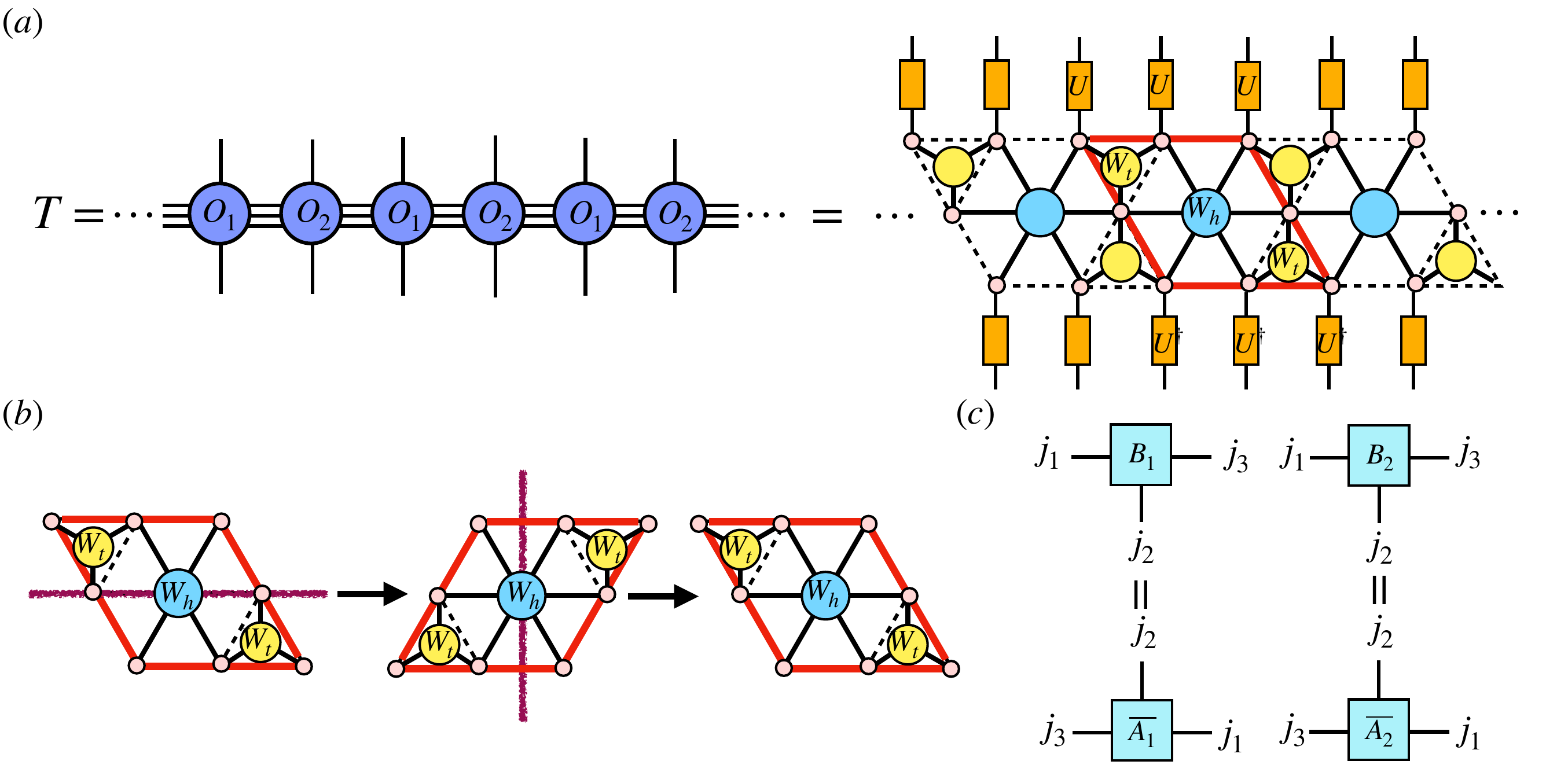}
\caption{(a) The tensor network representation of the transfer matrix, where the basic unit is indicated by the red line. The tensors enclosed by the dashed lines represent a row of Boltzmann weights, while the orange $U$ tensors correspond to the Fourier transform, given by $U(n, \theta) = e^{-in\theta}$.
(b) The reflection operation on the parallelogram unit cell, with the reflection axis denoted by the pink line.
(c) The relationship between the local tensors of up and down leading eigenvectors.}
\label{fig:Vomps}
\end{figure}

The transfer matrix $T$ possesses a spatial symmetry inherited from the lattice symmetry of the model, which can be used to simplify the leading eigenequations.
As illustrated in Fig.~\ref{fig:Vomps}(a), the row-to-row transfer matrix can be formally interpreted as the Fourier transformation of a row of Boltzmann weight:
\begin{multline}
    T^{\dots, n_1,n_2,n_3,\dots}_{\dots,m_1,m_2,m_3,\dots}=\prod_{i\in r}\int\frac{d\theta_i}{2\pi}\prod_{j}U(n_j,\theta_j)\\
    \cdot\prod_{p\in r}W_p(\{\theta_p\})\prod_{k\in r}U^{\dagger}(\theta_k,m_k).
    \label{eqn:transfer_matrix}
\end{multline}
Here, ${n_j}$ and ${m_k}$ represent the up and down indices of the transfer matrix, and $p$ denotes the translational invariant unit, which is a parallelogram region containing two triangle plaquettes and one hexagonal plaquette. The $\prod_{p\in r}W_p(\{\theta_p\})$ term represents a row of Boltzmann weights and can be regarded as a linear transfer matrix with continuous degrees of freedom. Since $W_p(\{\theta_p\})$ is real, invariant under the transformation:$\{\theta_p\}\rightarrow \{-\theta_p\}$, but not invariant under up-down transpose, the transfer matrix is invariant under complex conjugation and has a parity-time reversal (PT) symmetry \cite{Meisinger_1989}, but is non-hermitian.

The parallelogram region and the Boltzmann weight $W_p(\{\theta_p\})$ on the region remain unchanged under the combination of up-down reflection and left-right reflection, as depicted in Fig.\ref{fig:Vomps}(b). Consequently, the transfer matrix $T$ represented by Eq.(\ref{eqn:transfer_matrix}) is invariant under the joint operation of hermitian conjugation and left-right reflection.

At the level of local tensors, by utilizing the spatial symmetry, we only need to solve the leading down eigenvector $|\psi(A_1,A_2)\rangle$ and obtain the up eigenvectors from the leading bottom eigenvector:
\begin{align}
    &B_1(j_1,j_2,j_3)=\overline{A_1}(j_3,j_2,j_1)\notag\\&B_2(j_1,j_2,j_3)=\overline{A_2}(j_3,j_2,j_1),
    \label{eqn: relation between up and down eigenvectors}
\end{align}
where the bar denotes the complex conjugation of the elements of the tensors.

As the transfer matrix is non-hermitian, the fixed point equation of the transfer matrix can be solved by VOMPS algorithms \cite{Laurens_2019}, which is a combination of the power method and variational principle. The approximate leading vector is obtained by starting from a random uniform matrix product state and applying the transfer matrix to MPS repeatedly. In each power method step, the new MPS $T(k_1,k_2)|\psi(A_1,A_2)\rangle$ with bond dimension $dD$ is truncated to an MPS $|\psi( A_1^{'},A_2^{'})\rangle $ with bond dimension $D$ by maximizing the fidelity between the two states:
\begin{equation}
    \max_{A_1^{'},A_2^{'}}\frac{|\langle \psi(\overline{A_1^{'}},\overline{A_2^{'}})|T(k_1,k_2)|\psi(A_1,A_2)\rangle|^2}{\langle \psi(\overline{A_1^{'}},\overline{A_2^{'}})|\psi(A_1^{'},A_2^{'})\rangle}.
\end{equation}
This optimization problem for tensor $A^{'}$ can be accurately solved using the tangent-space method for uniform MPS \cite{Laurens_2019}. Surprisingly, although the thermal phase below the BKT transition temperature is a gapless phase, we find that the VUMPS algorithm can still find converged solutions, which are consistent with the solutions obtained by the VOMPS algorithm. The role of spatial symmetry of the transfer matrix in stabilizing the TN contraction algorithm has been discussed in several works \cite{Nyckees_2023,Tang_2023} and remains to be further explored.

\subsection{Calculation of generalized entanglement entropy}

\begin{figure*}[t]
\centering
\includegraphics[width=\linewidth]{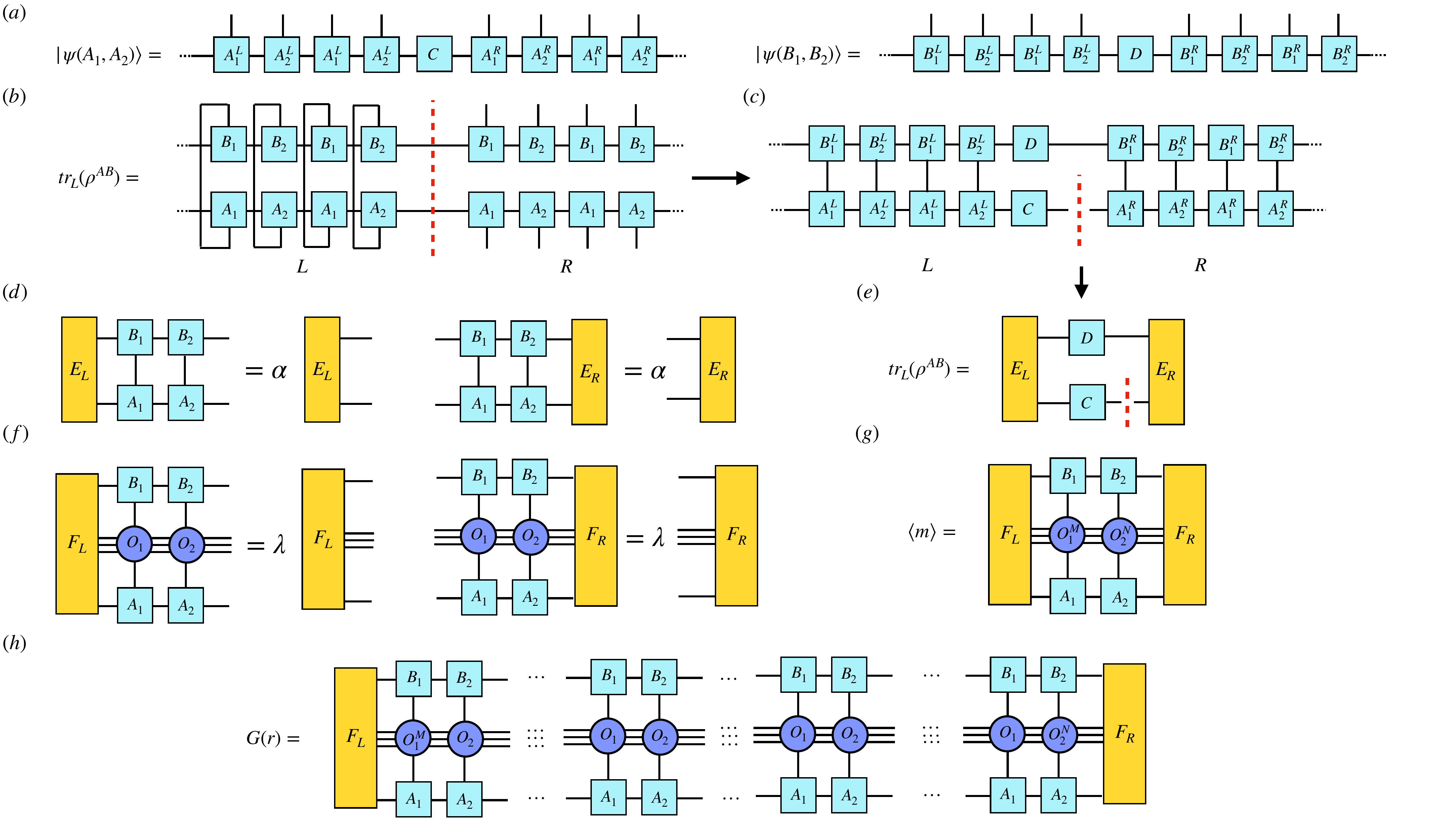}
\caption{Tensor network diagrams of physical quantities.
(a) The mixed canonical form of the uniform matrix product state.
(b) TN representation of the reduced density matrix. The system is divided into left and right parts by the dashed line, with the physical degrees of freedom in the left part traced out.
(c) Transformation of the reduced density matrix into a matrix with dimension $D\times D$ by changing the basis.
(d) The left and right leading eigenequations for the MPS transfer matrix.
(e) Calculation of the reduced density matrix by contracting the leading eigenvectors of the MPS transfer matrix.
(f) The left and right leading eigenequations for the channel operators.
(g) TN diagram of the expectation of a local operator.
(h) TN diagram of the two-point correlation function.}
\label{fig:EE_calculation}
\end{figure*}

For a hermitian transfer matrix, the entanglement properties of the system are encoded in the density matrix constructed by the leading eigenvector $\rho_A=|\psi(\overline{A})\rangle \langle\psi(A)| $ and the bipartite entanglement entropy $S_E$ can be calculated by the Schmidt decomposition of $|\psi(A)\rangle$, as illustrated in Fig.\ref{fig:EE_calculation}(a):
\begin{equation}
|\psi(A)\rangle=\sum_{\alpha,\beta=1}^{D} C_{\alpha,\beta}|\psi^L_{\alpha}\rangle|\psi^R_{\beta}\rangle,\quad S_E=-\text{Tr}(C^2\ln C^2),
\end{equation}
where $L$ and $R$ represent the two parts of the system.

This concept is generalized to the non-hermitian system \cite{Yi-Ting_2022, Tang_2023}. As shown in Fig. \ref{fig:EE_calculation}(b), the reduced density matrix of the non-hermitian transfer matrix $T$ is defined by
\begin{equation}
    \text{Tr}_L(\rho^{AB})=\frac{1}{\langle \psi^B|\psi^A\rangle}\sum_{\{\alpha_L\}}\langle \alpha_L|\psi^A\rangle\langle \psi^B|\alpha_L\rangle,
    \label{eqn: reduced_dm}
\end{equation}
 where $|\psi_A\rangle$, $\langle\psi_B|$ are the up and down leading eigenvectors, $\{\alpha_L\}$ is the complete orthogonal basis of the left part of the system.
 By using the mixed canonical form of the uniform matrix product state:
\begin{eqnarray}
&&|\psi(A)\rangle=\sum_{\alpha,\beta=1}^{D} C_{\alpha,\beta}|\psi^L_{A,\alpha}\rangle|\psi^R_{A,\beta}\rangle, \\&&|\psi(B)\rangle=\sum_{\alpha,\beta=1}^{D} D_{\alpha,\beta}|\psi^L_{B,\alpha}\rangle|\psi^R_{B,\beta}\rangle,
\end{eqnarray}
where $\langle \psi^x_{y,\alpha}|\psi^x_{y,\beta}\rangle=\delta_{\alpha,\beta}, x\in\{L,R\}, y\in \{A,B\}$, the matrix elements of $Tr_L(\rho^{AB})$ in the basis of $\{|\psi_{B,\beta}^R\rangle\}$ can be expressed as:
\begin{eqnarray}
    &&\langle \psi_{B,\alpha}^R|\text{Tr}_L(\rho^{AB})|\psi_{B,\beta}^R\rangle\notag\\
    &&=\sum_{\{\alpha_L\}}\sum_{i,j}C_{ii}D_{jj}\langle \alpha_L|\psi^L_{A,i}\rangle\langle \psi^L_{B,j}|\alpha_L\rangle\times \notag\\
    &&\langle \psi_{B,\alpha}^R|\psi_{A,i}^R\rangle\langle \psi_{B,j}^R|\psi_{B,\beta}^R\rangle\notag\\
    &&=\sum_{i,j}C_{ii}D_{jj}\langle \psi^L_{B,j}| (\sum_{\{\alpha_L\}}|\alpha_L\rangle\langle \alpha_L|)|\psi^L_{A,i}\rangle\times \notag\\
    &&\langle \psi_{B,\alpha}^R|\psi_{A,i}^R\rangle\delta_{j\beta}\notag\\
    &&=\sum_{i}C_{ii}D_{\beta\beta} \langle \psi^L_{B,\beta}|\psi^L_{A,i}\rangle\langle \psi_{B,\alpha}^R|\psi_{A,i}^R\rangle
\end{eqnarray}
   The corresponding TN diagram is plotted in Fig.~\ref{fig:EE_calculation}(c). The inner product
$\langle \psi^L_{B,\beta}|\psi^L_{A,i}\rangle$ and $\langle \psi_{B,\alpha}^R|\psi_{A,i}^R\rangle$ can be further squeezed into a small network by solving the left and right eigenvectors of the transfer matrix of MPS, as is displayed in Fig.\ref{fig:EE_calculation}(d). The final TN expression of the reduced density matrix is shown in Fig.\ref{fig:EE_calculation}(e). The generalized entanglement spectrum $\{s_{\alpha}\}$ can be calculated by diagonalizing the matrix in Fig.\ref{fig:EE_calculation}(e)
and the generalized entanglement entropy is determined by
\begin{equation}
    S_E=-\sum_{s\in\{s_\alpha\}}|s|^2\ln |s|^2
    \label{eqn:generalized_EE}
\end{equation}

\subsection{Calculation of local observable and correlation function}
In the $U(1)$ symmetric system, the local observable we are interested in at a certain site $O$ is of the form of $\langle e^{im\theta_o}\rangle$
\begin{equation}
    \langle e^{im\theta_o}\rangle=\frac{1}{Z}\prod_{i}\int\frac{d\theta_i}{2\pi}e^{-\beta E(\{\theta_i\})}e^{im\theta_o},
\end{equation}
where $m$ can take integer values. Following the preceding construction procedure, this can also be represented as a TN with discrete indices. The only difference is, at the site O, the original $\delta$ tensor defined by Eq.(\ref{eqn: delta_tensor}) is changed into an impurity tensor $\delta^m$:
\begin{eqnarray}
    \delta^m(n_1,n_2,n_3,n_4)&=&\int\frac{d\theta_o}{2\pi}\prod_{i=1}^{4}U(n_i,\theta_o)e^{im\theta_0} \notag\\
    &=&\delta_{n_1+n_2+n_3+n_4+m,0}.
    \label{eqn:impurity tensor}
\end{eqnarray}
We can further split the $\delta^m$ tensor in the same manner as shown in the main text.
\begin{equation}
    \delta_{n_1+n_2+n_3+n_4+m,0}=\sum_{n_0}\delta_{n_1+n_2+m-n_0,0}^{1,m}\delta_{n_3+n_4+n_0,0}^{2}.
\end{equation}
Again by grouping the $\{\delta^1,\delta^{1,m},\delta^2,\delta,I_h,I_t\}$ tensors,
the impurity tensor can then be compressed into a single row of linear transfer matrix, without changing the transfer matrix in other rows.
\begin{equation}
    \langle e^{im\theta}\rangle=\frac{1}{Z}\text{tTr}(\dots TTT_mTTT\dots).
\end{equation}
Similarly, The two points observable $\langle e^{im\theta_i}e^{-im\theta_j}\rangle$ can be represented as a TN which the corresponding local tensors on $i$ and $j$ sites are replaced by the impurity tensors defined by Eq.(\ref{eqn:impurity tensor}).

The TN with impurity tensors can be further compressed into an infinite train of local tensors $\{O\}$ sandwiched between the up and down eigenvectors
\begin{equation}
    z=\langle \psi(B_1,B_2)|T(k_1,k_2)|\psi(A_1,A_2)\rangle,
    \label{eqn: sandwitch_structure}
\end{equation}
\begin{equation}
    \langle M_iM_j\rangle=\frac{1}{z}\langle \psi(B_1,B_2)|T_M(k_1,k_2)|\psi(A_1,A_2)\rangle
     \label{eqn: observable_rep}
\end{equation}
where $T_M$ denotes the transfer matrix that contains the impurity tensors.
Here we assume the MPS is properly normalized as $\langle \psi(B_1,B_2)|\psi(A_1,A_2)\rangle=1$.

Eq.(\ref{eqn: sandwitch_structure}) and  Eq.(\ref{eqn: observable_rep}) can be further contracted by solving the leading eigenvalue of the channel operator.
For a given local tensor $O_i$, the channel operator is defined as
\begin{equation}
    C_{O_i}=\sum_{a,b} B^{b}_i\otimes O_i^{b,a}\otimes A_i^{a},
\end{equation}
where $a,b$ are the indices of the physical bond, and $i$ is the label of positions in the TN.
The eigenequations are then given by
\begin{equation}
    \langle F_L| C_{O_1}C_{O_2}=\lambda \langle F_L|,\quad C_{O_1}C_{O_2}|F_R\rangle=\lambda |F_R\rangle,
\end{equation}
as depicted in Fig.~\ref{fig:EE_calculation}(f). Once the equation is solved, the free energy density of the system can be directly obtained by
\begin{equation}
    f=-\frac{1}{3\beta}\ln\lambda,
\end{equation}
where the coefficient is the number of original physical spins contained in the unit cell of the TN.

With the help of the left and right leading vectors $\langle F_L|$ and $|F_R\rangle$,
The expectation value of local observables $\langle M_i M_{i+1}\rangle$ and two points correlation function $G(r)=\langle M_i M_{i+r}\rangle$ can also be reduced to a finite network
\begin{equation}
\langle M_i M_{i+1}\rangle= \langle F_L|C_{O_i}C_{O_{i+1}}| F_R\rangle,
\end{equation}
\begin{equation}
    G(r)=\langle F_L|C_{M_i}C_{O_{i+1}}C_{O_{i+2}}\dots C_{O_{i+r-1}}C_{M_{i+r}}|F_R\rangle.
\end{equation}
as shown in Fig.~\ref{fig:EE_calculation}(g) and (h).

\section{Numerical data}
\label{sec: data}
\subsection{Thermal melting of the vortex Lattice}

\begin{figure}[t]
\centering
\includegraphics[width=\linewidth]{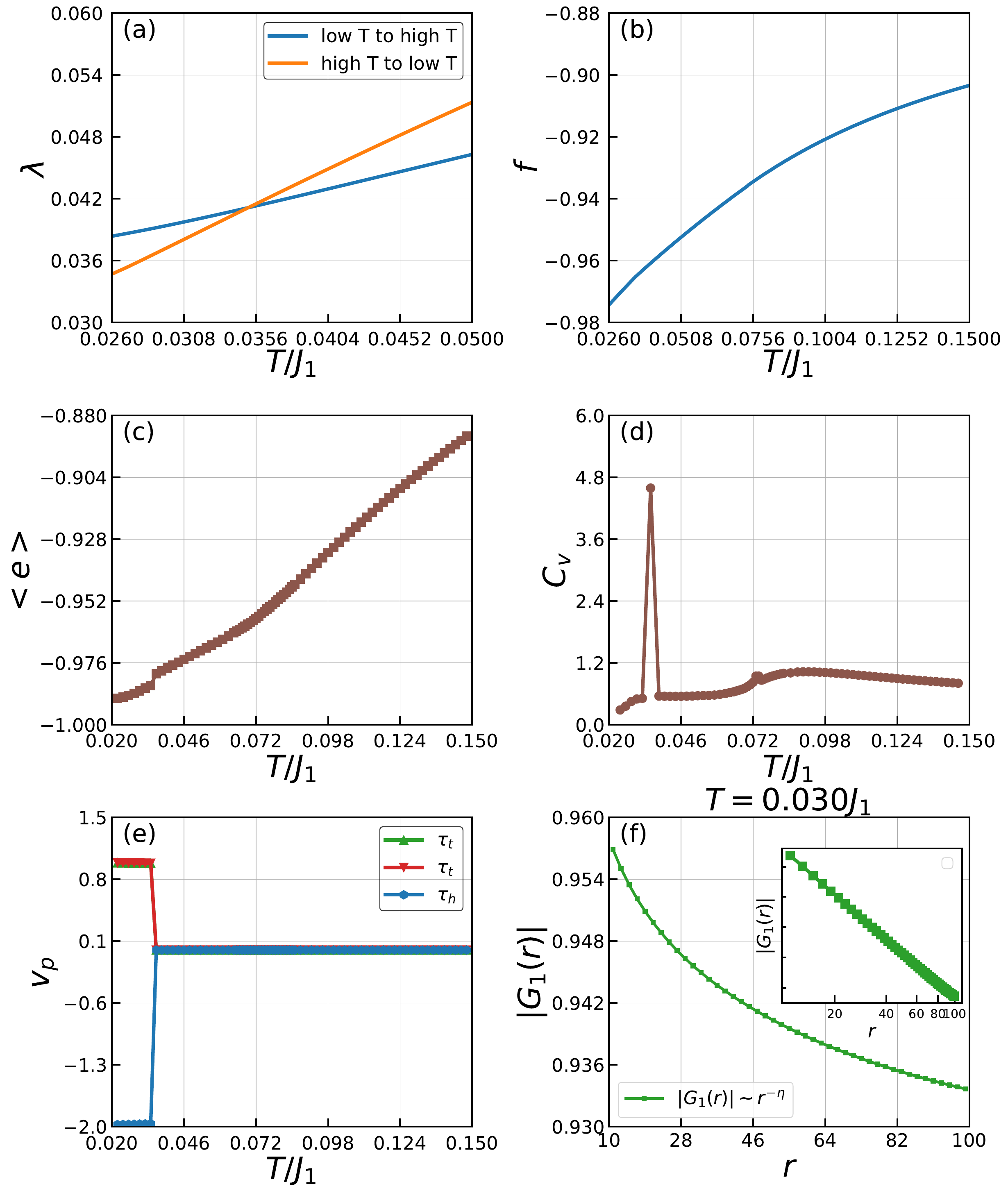}
\caption{Thermal quantities at $J_2=0.0025J_1$. (a) The leading eigenvalue of the two distinct solutions around the transition temperature $T_{c1}$. (b) The free energy as a function of temperature. (c) The internal energy as a function of temperature. (d) The specific heat as a function of temperature. (e) The chirality as a function of temperature. (f) The amplitude of the correlation function of the integer vortices at $T = 0.030J_1$. Inset: The correlation function of the integer vortices exhibits a power-law decay.}
\label{fig:thermal_quantities}
\end{figure}

We further discuss the thermal properties of the $J_1$-$J_2$ AFXY model on a kagome lattice in greater detail. Near the first-order transition $T_{c1}$, two distinct solutions emerge depending on the initial state. As shown in Fig.~\ref{fig:thermal_quantities}(a), the leading eigenvalue $\lambda$ exhibits thermal hysteresis, a clear hallmark of a first-order transition \cite{Song_2023_2}. The precise location of the first-order transition can be determined by the intersection of the two solutions, with the physical result corresponding to the solution with the larger leading eigenvalue or the lower free energy density.

In Figs.\ref{fig:thermal_quantities}(b)-(d), we present the temperature dependence of the free energy density, internal energy density, and specific heat, respectively. The internal energy density of the system is calculated by summing the expectation value of nearest-neighbor two-body interaction terms and next-nearest-neighbor two-body interaction terms in a unit cell. The specific heat can be calculated by numerically differentiating the internal energy density $u(T)$, $C_v=\frac{du}{dT}$. The specific heat data exhibit a sharp singularity at $T_{c1}$, consistent with the first-order transition. At $T_{c2}$, the internal energy density shows no singular behavior, and the specific heat develops a small rounded bump above $T_{c2}$, which is a characteristic feature of a BKT transition.

The low-temperature phase below $T_{c1}$ can be characterized by the chirality on each plaquette and the behavior of the spin-spin correlation function. The chirality is calculated as follows:
\begin{equation}
\tau_p=\sum_{\langle i,j\rangle\in p}\frac{2}{3\sqrt{3}}\langle \sin(\theta_i-\theta_j)\rangle,
\label{eqn:chirality}
\end{equation}
where $\frac{2}{3\sqrt{3}}$ is a normalized coefficient. This quantity has the same physical meaning as vorticity and is more convenient to calculate within the framework of TN, as each term in Eq.(\ref{eqn:chirality}) can be obtained from the imaginary part of $\langle e^{i(\theta_i-\theta_j)}\rangle$. Below the first-order transition line, as displayed in Fig.\ref{fig:thermal_quantities}(e), the chirality $\tau_t$ on both the up-type triangle and the down-type triangle takes a value close to $1$. Meanwhile, the chirality $\tau_h$ on each hexagonal plaquette takes a value close to $-2$, which is in agreement with the vortex-antivortex lattice ground state of the model.

Moreover, the spin-spin correlation function is expressed as:
\begin{eqnarray}
&&G_1(r)=\langle e^{i(\theta_i-\theta_{i+r})}\rangle=|G_1(r)|e^{i\phi_1(r)},
\end{eqnarray}
where $G_1(r)$ characterizes the behavior of integer vortices. $|G_1(r)|$ at $J_2=0.0025 J_1$ and a given temperature $T=0.03 J_1< T_{c1}$ is shown in Fig. \ref{fig:thermal_quantities}(f). The amplitude of $G_1(r)$ can be well-fitted by a power-law function with distance:
\begin{equation}
|G_1(r)|\sim r^{-\eta_1},
\end{equation}
which suggests that the integer vortex and anti-vortex are bound into small pairs.
Above the first-order transition line, $G_1(r)$ shows an exponential decay behavior, and the vorticities on the plaquettes jump discontinuously to zero.

Compared to previous theoretical analyses \cite{Korshunov_2002}, we find a broader region for the fractional vortex pair phase ($J_2/J_1<0.005$) than the theoretical estimation ($J_2/J_1<10^{-4}$). By combining the analysis of fractional vortex excitations with numerical results, it can be observed that the fractional vortex paired phase only exists when $E_{dw}<10^{-2}E_{v}$. For a general frustrated XY model with both continuous and discrete degeneracy, the energy of the domain wall and the pre-logarithmic factor of vortex energy are of the same order \cite{Korshunov_2006}. The condition $E_{dw}<10^{-2}E_{v}$ is likely to occur only under the circumstance of extensive ground state degeneracy in the system.

\bibliography{ref}

\end{document}